%% file: main.tex
\documentclass[a4paper,12pt]{article}

\input{mystyle.sty}

\usepackage{geometry}
\usepackage{todonotes}

\begin{document}

\title{Beyond Barker: Infant Mortality at Birth \\
and Ischaemic Heart Disease in Older Age}
\author{Samuel Baker\thanks{University of Bristol. E-mail: \href{mailto:samuel.baker@bristol.ac.uk}{samuel.baker@bristol.ac.uk}.} 
\and 
Pietro Biroli\thanks{University of Bologna and University of Z\"urich. E-mail: \href{mailto:pietro.biroli@unibo.it}{pietro.biroli@unibo.it}.} 
\and 
Hans van Kippersluis\thanks{Erasmus University Rotterdam. E-mail:  \href{mailto:hvankippersluis@ese.eur.nl}{hvankippersluis@ese.eur.nl}.} 
\and Stephanie von Hinke\thanks{Corresponding author: University of Bristol; Erasmus University Rotterdam. E-mail: \href{mailto:S.vonHinke@bristol.ac.uk}{s.vonhinke@bristol.ac.uk}. 
We would like to thank Alice Carter, seminar participants at the VU University Amsterdam, the Centre for Health Economics, and the Alpine Population Conference for valuable comments. We gratefully acknowledge employees and participants of the 23andMe, Inc cohort for sharing GWAS summary statistics for educational attainment. This work is based on data provided through www.VisionofBritain.org.uk and uses historical material which is copyright of the Great Britain Historical GIS Project and the University of Portsmouth. We gratefully acknowledge financial support from NORFACE DIAL (462-16-100) and the Medical Research Council (MR/N0137941/1). The project DONNI has also received funding from the European Research Council (ERC) under the European Union's Horizon 2020 research and innovation programme (grant agreement No. 851725).}}

\date{\today}
\maketitle

\begin{abstract}
\noindent \singlespacing
Adverse conditions in early life can have consequential impacts on individuals' health in older age. In one of the first papers on this topic, \cite{BarkerOsmond1986} show a strong positive relationship between infant mortality rates in the 1920s and ischaemic heart disease in the 1970s. 
We go `beyond Barker’, first by showing that this relationship is robust to the inclusion of local geographic area fixed effects, but \textit{not} family fixed effects. 
Second, we explore whether the average effects conceal underlying heterogeneity: we examine if the infant mortality effect offsets or reinforces one’s genetic predisposition for heart disease. We find considerable heterogeneity that is robust to within-area as well as within-family analyses. Our findings show that the effects of one's early life environments mainly affect individuals with the highest genetic risk for developing heart disease. Put differently, in areas with the lowest infant mortality rates, the effect of one’s genetic predisposition effectively vanishes. These findings suggests that advantageous environments can cushion one's genetic risk of developing heart disease. 
\vspace{2mm}
\end{abstract}

\newpage
\onehalfspacing


\doublespacing
\section{Introduction}
\label{sec:intro}
Adverse conditions during the prenatal and early childhood period can have significant and potentially irreversible impacts on individuals' health and well-being in older age \citep[for a recent review, see e.g.][]{Gluckman2016,Gage2016,Conti2019}.\footnote{The vast literature on the so-called Developmental Origins of Health and Disease (DOHaD) hypothesis spans both the medical and social sciences \citep[see e.g.,][]{Gillman2005,GluckmanEtAl2008, AlmondCurrie2011b, AlmondCurrie2011a,almond2018childhood,Heindel2017,Arima2020}. Correlational and causal linkages have been found between several measures of early life adversities and health outcomes---such as cardiovascular disease, metabolic syndrome, and obesity---as well as socio-economic outcomes---such as educational attainment and wages.}  
The best known proponent of this hypothesis is the British physician and epidemiologist David Barker. In one of the first of a set of papers, Barker and colleagues showed a strong positive geographical relationship between the infant mortality rate in the 1920s and ischaemic heart disease mortality in the 1970s in the UK \citep{BarkerOsmond1986}. They conclude that ``poor nutrition in early life increases susceptibility to the effects of an affluent diet.''\footnote{Some of his other papers on this topic include, e.g. \cite{Barker1987}, focusing on maternal mortality as a proxy for the early life environment, and \cite{barker1989intrauterine}, where early life neonatal and postneonatal mortality predict later life mortality from stroke. In other work, they show that low birth weight associates with mortality from ischaemic heart disease in older age \citep{barker1989weight}.}

In addition to such `environmental' circumstances affecting the development of heart disease, genetic factors are known to play an important role. For example, twin studies have shown that the disease is heritable, and more recent Genome-Wide Association Studies (GWAS) have started to unravel the specific genetic variants implied in the disease \citep[see e.g.][]{samani2007genomewide,helgadottir2007common,mcpherson2007common,schunkert2011large,deloukas2013large,nikpay2015comprehensive,howson2017fifteen,nelson2017association,van2018identification}.
These gene-discovery studies have linked dozens of independent genetic loci to heart disease, facilitated a better understanding of the causal risk factors, and informed the development of new therapeutics \citep[see e.g.][]{khera2017genetics}. 

Whereas the role of these environmental and genetic main effects are established and widely-documented, the understanding of their interplay---so-called Gene-Environment (GxE) interplay---is still in its infancy.\footnote{Some studies discuss the potential role of gene-environment interactions for cardiovascular disease, but none of these focus on the long-held DOHaD hypothesis \citep[see e.g.,][]{Visvikis2008,Hirvonen2009,Flowers2012}.} 

In this paper, we start by laying out different mechanisms through which genetic variation can moderate the effect of environmental conditions. We highlight the potential role of epigenetics, but argue that this is neither a \textit{sufficient}, nor a \textit{necessary} condition for the existence of gene-environment interaction effects. We then empirically confirm the independent importance of genetic variation and adverse conditions during gestation and early childhood \citep[as proxied by infant mortality rates at birth as in][]{BarkerOsmond1986}, in explaining later life heart disease. We then go `beyond Barker' in two ways. First, we explore the robustness of the Barker hypothesis to the inclusion of local geographical area fixed effects, exploiting only variation in infant mortality rates and heart disease \textit{within} a local area over time. We also investigate the sensitivity to the inclusion of family fixed effects, comparing later life heart disease among full siblings who were exposed to different rates of infant mortality at birth, whilst controlling for individuals' age at the \textit{monthly} level. 
Second, we move `beyond Barker' by studying the importance of gene-environment interplay in ischaemic heart disease. This allows us to answer questions such as ``does genetic susceptibility aggravate adverse early life circumstances?'', or -- vice versa -- ``can advantageous early life environments cushion genetic risk?'' 

We do this using individual-level data on over 370,000 individuals in the UK Biobank \citep{Sudlow2015}, of whom we identify $\sim$33,000 full siblings from their genetic data.
Ischaemic heart disease is the most common cause of death in the developed world, accounting for more than nine million deaths worldwide in 2016 \citep[e.g.,][]{WHO2018}.\footnote{Ischaemic heart disease occurs when the arteries cannot transfer enough oxygen-rich blood to the heart. The most common type is coronary artery disease, where the blood-flow to the heart is restricted by atherosclerosis, or plague.} 
Advancing our knowledge on the interplay between genetic and environmental factors that drive the world's major killer is therefore not just a fundamental scientific advance, but may also inform governments on how environmental policies can reduce the -- arguably unfair -- inequalities in heart disease arising from one's genetic variation. 

Our first contribution is to digitize historical data on infant mortality rates in the UK\footnote{We start with the already-digitized data from the Great Britain Historical Database \citep[GBHD;][]{GBHD2020}, see also the \hyperlink{https://www.visionofbritain.org.uk/}{Vision of Britain} website. The GBHD contains birth, death, and infant mortality counts, as well as population estimates from 1930 to 1973 but lacked the years 1958 to 1962. We digitized these remaining years, and systematically quality-controlled the entirety of the database.}, link these to the UK Biobank and use them to replicate the analysis by \cite{BarkerOsmond1986} on the relationship between early life conditions and later life outcomes.  
Rather than specifying \textit{area}-level outcomes, as in the original study, we have \textit{individual}-level outcomes for those born between the late 1930s and early 1970s. 
As a measure of the early life environment, we use the infant mortality rate that each individual was exposed to in their year and local area of birth, as published in the Registrar General reports for England and Wales \citep{RegistrarGeneralReports}.
The spatial units we employ are Local Government Districts, including over 1400 small geographic regions covering the whole of England and Wales. 
We then merge this information with the UK Biobank, a major resource that follows the health and well-being of around 5 million individuals in the UK aged 40-69 between 2006-2010. 
Linking these two datasets provides previously unexplored information on the early life conditions in the $\sim$35 years and $\sim$1400 districts of birth of the UK Biobank participants. This in turn allows us to explore the determinants of heart disease, combining spatial variation in infant mortality rates across small geographical regions \citep[\`a la][]{BarkerOsmond1986} as well as temporal variation in infant mortality rates across individuals' year of birth \citep[\`a la][]{Kermack1934}. Indeed \citet{Kermack1934} were among the first to suggest that one's childhood environment is crucial in shaping individuals' health in adulthood, but they largely ignore spatial variation in childhood environments. In contrast, \citet{BarkerOsmond1986} focus on such geographical variation, but do not explore temporal variation in adverse environments. Our data allow us to combine both sources of variation and investigate the relationship between early life adversity and later life health for a period characterised by a substantial improvement in early life conditions. 

A second contribution is to deepen our understanding of the relationship between early life environments and later life outcomes by incorporating recent advances in the understanding and collection of individual-level molecular genetic data. 
More specifically, we investigate whether the early life environment exacerbates or mitigates any genetic effects on heart disease. We begin by showing the strong predictive power of the early life environment as well as the relevant polygenic score\footnote{Polygenic scores measure one's genetic `predisposition' towards a trait (here: heart disease). We discuss this in more detail below.}, confirming the importance of both components. Next, we investigate the interaction between the polygenic score and the infant mortality rate at the time and location of birth. 
In other words, we shed light not only on whether improving the early life environment reduces the \textit{average} disease prevalence in older age, as suggested by Barker, but also whether such improvements reduce the \textit{inequality} in disease prevalence for individuals with different genetic predispositions for the disease. 
In the other direction, our interaction results enable testing whether individuals' genetic predisposition exacerbates or protects individuals against adverse early life conditions in developing ischaemic heart disease.

We have three main findings. 
First, we replicate Barker's result that the infant mortality rate in one's local district and year of birth is associated with ischaemic heart disease later in life using well-powered individual-level data from the UK Biobank. However, our interpretation of these findings differs from Barker on two counts. First, we show that infant mortality rates are systematically associated to socioeconomic disadvantage, and hence are likely to additionally capture other characteristics of the environment than just ``nutritional deficiencies'' \citep[as in][]{BarkerOsmond1986}. Indeed, areas with high infant mortality rates are poorer and of lower socioeconomic position, and this in itself may negatively affect individuals' health in later life. Second, we explore the robustness of our findings to the inclusion of Local Government District fixed effects, as well as to family fixed effects. In the former, we compare people born in the same district over time and exploit variation in infant mortality rates and outcomes \textit{within} districts.
In the latter, we compare siblings born in different years and exploit variation in infant mortality rates and outcomes \textit{within} families. 
We find that the main effect of the infant mortality rate is robust (albeit attenuated) to accounting for time-invariant variation within districts, but \textit{not} within families. This suggests that the infant mortality rate does not mainly capture within-family variation in nutritional circumstances, but rather that it proxies for other unobserved characteristics that vary between families, such as poverty or socio-economic status. Accounting for these using a family fixed effects approach renders the average effect of the infant mortality rate insignificant. 

A second main finding is that the genetic signal in ischaemic heart disease largely derives from direct (or causal) genetic effects. As pointed out by \cite{DaveySmith2003,howe2021within,Biroli2022} among others, family designs enable us to estimate the direct genetic effect on the outcome, accounting for demography (e.g. population stratification, assortative mating) and indirect genetic effects (e.g., genetic nurture, where parental genotypes influence offspring outcomes through environmental channels). We find that the genetic effects are robust to the inclusion of district and family fixed effects, suggesting that demography and indirect genetic effects are not driving the association between polygenic scores and heart disease. To clarify: by direct genetic effects we make a probabilistic statement about a counterfactual, i.e., those with higher polygenic scores are \textit{more likely} to develop ischaemic heart disease later in life. We do \textit{not} mean that these effects are purely biological, immutable or deterministic. Indeed, even though these effects are downstream consequences of genetic variation across individuals, they could be entirely mediated by malleable environmental conditions. 

The third main finding is that the null effect of infant mortality conceals underlying genetic heterogeneity: we find evidence of non-negligible interactions between genes and the environment. Our findings imply that among those with high genetic risk for ischaemic heart disease, the infant mortality rate significantly aggravates the risk. In contrast, in districts with the lowest infant mortality rates, the effect of genetic predisposition effectively vanishes. We find that these results are robust (albeit attenuated) to the inclusion of district as well as family fixed effects.
This not only reveals previously masked heterogeneity in the DOHaD literature, but additionally provides evidence that environmental interventions may moderate and even mitigate genetic susceptibility to heart disease.\footnote{Although we know from the epigenetics literature that early-life circumstances may trigger differential epigenetic expression \citep[see e.g.,][]{heijmans2008persistent}, much remains to be learned. First, epigenetic studies generally do not investigate how epigenetic changes affect later life outcomes. Instead, they tend to explore how different circumstances, or environments, affect DNA methylation. Second, not all epigenetic effects are expected to translate into developmental differences, and they could arise in segments of the DNA where there is no variation across individuals. In fact, as we argue below, epigenetics is one potential channel through which gene-environment interplay exists, but it is neither necessary, nor sufficient.} 
This finding strongly debunks genetic determinism of later life diseases such as heart disease, and establishes that the disease is the product of complex interactions between genes and the environment. 
As such, it suggests that improving the early life environment can significantly reduce the variation driven by `genetic risk' in the population.

The rest of this paper is structured as follows: \hyperref[sec:background]{Section~\ref*{sec:background}} provides a background to the paper, discussing the importance of `nature' (genetics), `nurture' (proxied here by the infant mortality rate) and why or how the two may interact to shape individuals' outcomes later in life. The data is described in \hyperref[sec:data]{Section~\ref*{sec:data}}, followed by the empirical specification in \hyperref[sec:methods]{Section~\ref*{sec:methods}}. The results are discussed in \hyperref[sec:results]{Section~\ref*{sec:results}}, with the robustness analysis in \hyperref[sec:robustness]{Section~\ref*{sec:robustness}}. We conclude in \hyperref[sec:concl]{Section~\ref*{sec:concl}}.

\section{Background}\label{sec:background}

\subsection{Nature: Genetics}
\label{sec:G}

The human genome consists of over 3 billion base pairs in each cell nucleus, with four possible bases: adenine (A), thymine (T), guanine (G) and cytosine (C).\footnote{A base \textit{pair} is set of two bases, with A always pairing with T, and C always pairing with G.} Comparing any two unrelated human beings, over 99\% of their genome is identical. The remaining $<$1\% differs between individuals, with a Single Nucleotide Polymorphism (or SNP, pronounced \textit{snip}) being the most common form of genetic variation. A SNP is a one base-pair substitution at a particular location (locus) on the human genome.

To identify genetic variants that are associated with a particular trait of interest, such as coronary heart disease, so-called Genome-Wide Association Studies (GWAS) relate each SNP to the trait in a hypothesis free-approach. Stringent \textit{p}-values are then used to identify SNPs that are robustly associated with the trait of interest, and replication is performed in other, independent samples. Most human complex traits are polygenic, meaning that they are affected by many SNPs, each with a very small effect size. To increase the predictive power of SNPs, they can be aggregated into a so-called polygenic score (PGS, also referred to as a polygenic index or PGI), defined as a weighted sum of the individual SNPs:

\begin{equation*}
PGS_i = \sum_{j=1}^J \beta_j x_{ij}
\end{equation*}

where $x_{ij}$ is a count of the number of `effect' alleles at SNP $j$ of individual $i$, and $\beta_j$ is the associated effect size, obtained from an independent GWAS. 
Polygenic scores have been shown to be powerful tools to identify patients with increased risk for coronary artery disease, atrial fibrillation, type 2 diabetes, inflammatory bowel disease, and breast cancer. Indeed, \cite{khera2017genetics} propose the use of polygenic prediction in clinical care.

\subsection{Nurture: The infant mortality environment}
\label{sec:E}

The infant mortality rate in the year and location of birth can be seen as a broad measure of the environment that an individual was exposed to around birth. Different mechanisms have been suggested that explain how such early life environments can affect outcomes in older age. First, adverse conditions in early life may lead to permanent changes in body structure, physiology and metabolism \citep[see e.g.][]{GluckmanEtAl2008,belsky2019early, colich2020biological,McDermotte2105304118}. For instance, they may lead to compensatory patterns of growth, with nutritional resources being diverted from child development towards survival, potentially affecting the development of the body more generally. 
Similarly, adverse circumstances may cause the body to automatically protect the growth of one organ (e.g., the brain) at the expense of other organs (e.g., the heart), which in turn can affect the development of future disease \citep[see e.g.][]{CampbellEtAl1967, Rudolph1984,hales2001thrifty}.

Another mechanism is `foetal programming'. This refers to the intrauterine environment giving the foetus a forecast of the circumstances into which it will be born \citep{GluckmanHanson2006a,GluckmanEtAl2008}. Individuals' metabolism may then adapt to ensure optimal survival under similar conditions. Hence, foetal programming is believed to reflect foetal \textit{adaptation}. Whilst foetal adaptation may be beneficial to short-term survival, it could be detrimental to health in adulthood if later life circumstances are very different from those prenatally \citep{Barker1995}.
For example, if the foetus was exposed to a famine prenatally, the body may develop in a way to ensure survival under similar conditions post-birth. If it is then exposed to an environment with sufficient nutrition (or over-nutrition), this may impact on the development of disease later in life.\footnote{For studies looking at the effects of intrauterine exposure to famines on later life outcomes, see e.g. those on the Dutch Hunger Winter \citep{Smith1947,SteinEtAl1975,Schultz2010,conti2021severe}, famines in China \citep{st2005rates}, Finland \citep{kannisto1997no}, the Netherlands \citep{Lindeboom2010}, and Leningrad \citep{stanner1997does}.}

Finally, rather than the infant mortality rate having a causal effect on later life health, it is likely to also capture socio-economic differences. Indeed, the literature on infant mortality rates in the 20$^{th}$ century shows a strong social gradient: infant mortality is substantially higher among the unskilled social classes compared to the professional classes, and in northern compared to more southern regions within the UK. It is also higher among illegitimate children \citep{AdelsteinEtAl1980}.

\subsection{The potential interaction between `nature' and `nurture'}
\label{sec:GxE}

There are multiple reasons why `nature' might interact with `nurture' to shape individuals' later life outcomes. First, there may be a \textit{biological} channel: genes may predispose individuals to certain health conditions or behaviours, but the extent to which these genes are expressed -- their phenotypic effect -- can depend on environmental circumstances \citep{GluckmanHanson2006a,Rutter2006}. For example, if adverse circumstances early in life cause permanent changes in the structure of certain organs (e.g., the heart), genetic variants may change their expression in that organ. This can in turn affect the development of the organ and with that, future disease. This channel is also known as environmentally-induced epigenetic regulation \citep[see e.g.,][]{GluckmanHanson2006b}; switching genes on and off depending on environmental circumstances. The epigenetic literature tends to focus on the effects of certain exposures on the epigenome, with some studies establishing epigenetic expressions driven by early life circumstances \cite[e.g.,][]{heijmans2008persistent}.

However, even if early life circumstances cause epigenetic changes, this does not necessarily imply that they affect later life outcomes. 
Indeed, there could be compensatory investments by the individual or her environment that offset the development of any epigenetic effects on later life phenotypes. 
For example, parental investments in children may differ in response to early life circumstances, potentially mitigating or reinforcing their effects \citep[see e.g.][]{Yi2015,Adhvaryu2016ej,Gratz2016,Molina2021,Majid2018,Muslimova2020}. 
This may be a conscious reaction to environmental circumstances, but it could also be an unconscious response. Similarly, if prenatal malnutrition leads to epigenetic expression, but parents fully compensate for this harmful exposure, we would not necessarily detect any differences in later-life outcomes. 
In other words, epigenetic expression in response to environmental circumstances is not a \textit{sufficient} condition for the existence of gene-environment interaction effects on later life outcomes.

Furthermore, an interaction between nature and nurture need not be epigenetic. For example, one's genetic predisposition for certain health conditions or behaviours may simply reflect the `type' of individual, with different `types' responding differently to different environments. 
For example, \cite{Fletcher2012} finds that variation in the nicotinic acetylcholine receptor moderates the influence of tobacco taxation on smoking, showing that only those with the protective polymorphism respond to tobacco taxation and reduce their tobacco use \citep[see also, e.g.,][]{Slob2021}. A differential response to tobacco taxation is unlikely to occur due to epigenetic changes or compensatory effects, and is more likely to indicate that those with the protective polymorphism can be seen as different `types' who respond differently to different environments. In
other words, epigenetics is also not a \textit{necessary} condition for the existence of gene-environment interaction effects. 

In this study, we cannot identify the specific channel that may be driving our gene-by-environment interaction effects, and hence we are unable to distinguish between the different potential mechanisms. Instead, we quantify the extent to which genes and environments interact in shaping individuals' later life health, which is an essential precursor to analyses seeking to identify epigenetic changes or other channels through which the early-life environments and genetic predisposition interact in the development of heart disease.

\section{Data}
\label{sec:data}

The data we use is the UK Biobank; a major resource that follows the health and well-being of approximately 500,000 individuals in the UK aged 40-69 between 2006-2010. Participants have provided information on their health and well-being, and given blood, urine and saliva samples. They have also been genotyped.

An advantage of the UK Biobank is that it is a very large sample of individuals for whom we observe an extensive amount of relevant health (as well as social and economic) outcomes later in life. Our main outcome of interest is motivated by the existing literature: we define a binary indicator measuring whether the individual has been diagnosed with ischaemic heart disease (IHD). We identify individuals using the ICD-10 codes (I20-I25) obtained from mortality records as well as primary and secondary diagnoses in individuals' hospital inpatient records that are linked to the UK Biobank. Around 11\% of our sample has been diagnosed with IHD. 

Another advantage of these data is that it includes a relatively large sample of siblings. This means that the analysis can hold constant any time-invariant observed and unobserved family characteristics that may affect both the exposure and outcome of interest, using family fixed effects. We do this in the analysis below.

One of the main disadvantages of the UK Biobank, however, in particular for our research, is that there is almost no information on the early life environment, other than the location of birth.\footnote{The data also include (self-reported) birth weight, an indicator for whether the participants were breastfed, and whether the mother smoked during pregnancy.} To allow us to explore the effects of early life circumstances on later life health, we exploit the the eastings and northings of the location of birth to identify the Local Government District in which individuals were born.
We then merge in data on individuals' environmental conditions in their year of birth at the Local Government District level (henceforth: district-level). In the absence of information on individual- or family-level circumstances around birth, this allows us to \textit{characterise} the birth environment for each individual in the UK Biobank, providing a rich resource to add to these existing data.\footnote{Note that the districts we observe suffer from the modifiable areal unit problem \citep[MAUP,][]{Netrdova2020}, in that they change over time in terms of their shape, name and type (e.g., rural or urban district). A frequently-used solution is aggregation \citep{Gregory2005} 
to larger time-invariant geographic units, but this leads to a loss of data and specificity. An alternative is to create weights for smaller spatial sub-units (representing, e.g., its population) and use these to construct time-invariant districts and their associated (weighted) statistics.
We use WeightGIS \citep{Baker2020} to do the latter, creating time-invariant districts based on their parish-level population; we standardise districts to the 1951 census shapefile provided by Vision of Britain \citep{GBHDGIS2011}.}
  
To this end, we take the Great Britain Historical Database \citep[GBHD;][]{GBHD2020} as a starting point, which contains district-level birth, death, and infant mortality counts, as well as population estimates for the years 1930--1957 and 1963--1973. We collect and digitize this information for the remaining years 1958--1962, and systematically quality-control the entirety of the database. We then link the infant mortality rate in the year and district of birth, defined as the total number of deaths within the first year of life per 1000 live births, for all UK Biobank participants born in England or Wales. This local mortality rate is a proxy for the quality of the early life environment that each individual was exposed to. 
The year and district-specific infant mortality data are extracted from the Registrar General Statistical Reviews of England and Wales from 1934-1971 \citep{RegistrarGeneralReports}.

We use the genetic data in the UK Biobank to create a polygenic score for heart disease, using the GWAS summary statistics from \cite{nikpay2015comprehensive}. This GWAS excludes the UK Biobank, ensuring the discovery sample is independent from the analysis sample, meaning we do not suffer from over-prediction.\footnote{In the robustness analyses, where we focus on the sibling sample, we explore the sensitivity of our findings to the use of an alternative polygenic score, obtained from a GWAS on the UK Biobank sample that excludes siblings and related individuals. We then use these GWAS estimates to construct a polygenic score on the analysis sample of siblings. Our results are not sensitive to this alternative construction of the polygenic score.} We correct for co-inheritance between SNPs using LDpred \citep{vilhjalmsson2015modeling}, assuming the fraction of causal SNPs is 1. We standardise all polygenic scores to have mean 0, standard deviation 1.

We make the following sample selection: First, we drop those with missing birth co-ordinates since these cannot be geo-located to a Local Government District. We also drop those born outside England and Wales, as the infant mortality data does not contain reports for other parts of the United Kingdom. This leaves us with 405,180 individuals. We next drop those of non-European ancestry, those with missing data on ischaemic heart disease, or on infant mortality rates. Our final sample size includes 378,838 individuals. In further analysis, we restrict our estimation to the sibling sample of the UK Biobank. As there is no self-reported information on family members, we identify siblings using the (genetic) kinship matrix provided by the UK Biobank. Our analysis sample includes 33,060 full siblings. 


\subsection{The Infant Mortality Rate}
\label{sec:descr_IMR}

\begin{figure}[ht!]
\centering
\includegraphics[width=0.85\linewidth]{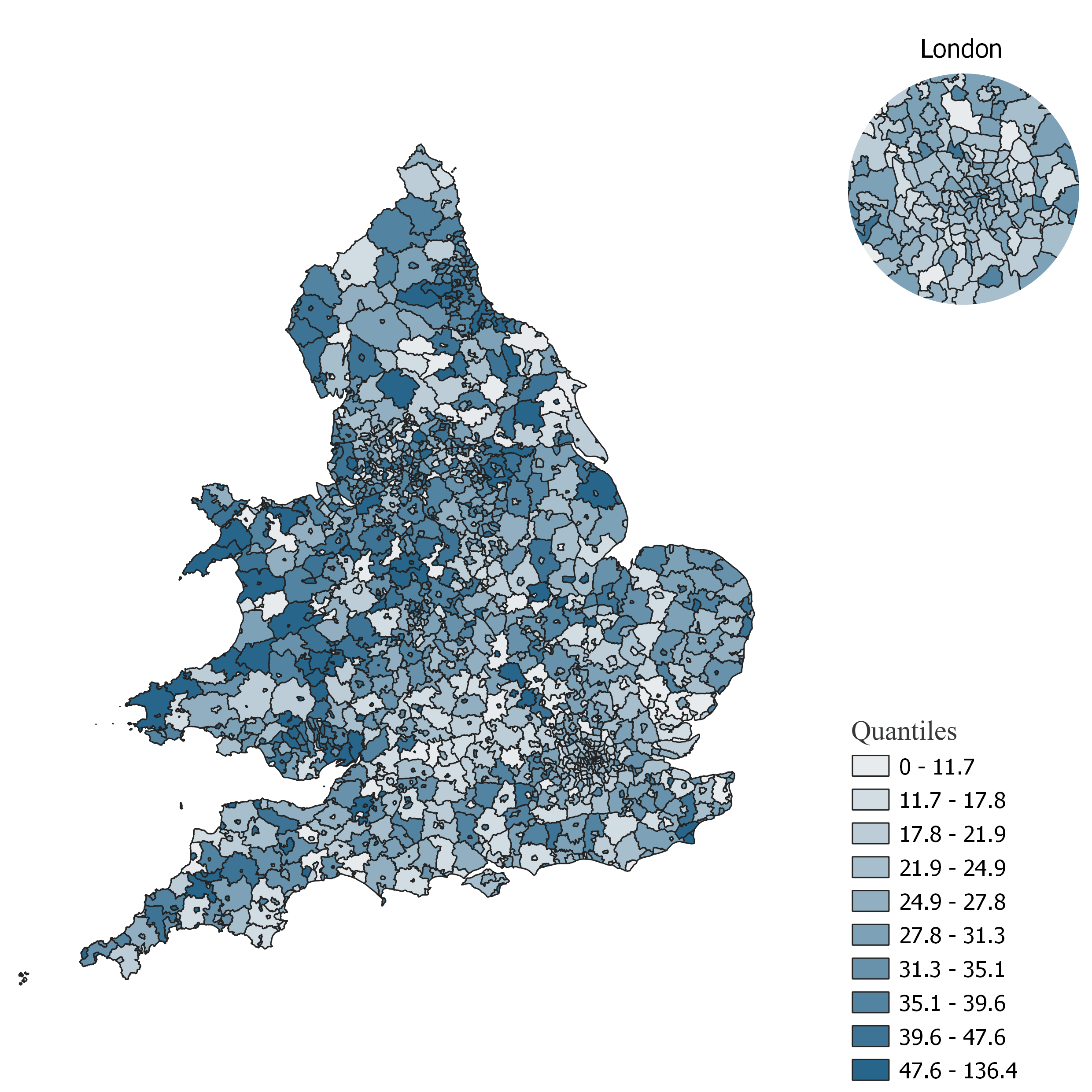}
\caption{Geographic distribution of the infant mortality rate (infant deaths per 1000 live births) in 1951 for England and Wales. This work is based on data provided through www.VisionofBritain.org.uk and uses historical material which is copyright of the Great Britain Historical GIS Project and the \citet{GBHDGIS2011}. }
\label{fig:InfantDeaths}
\end{figure}

\autoref{fig:InfantDeaths} shows the distribution of infant mortality rates across England and Wales for the 1951 Census year. The different geographical regions shown are the 1472 Local Government Districts for which we observe annual infant mortality rates.  These are relatively small geographic areas, showing substantial variation in infant mortality rates across England and Wales. 

\begin{figure}[ht]
  \centering
  \includegraphics[width=0.6\linewidth]{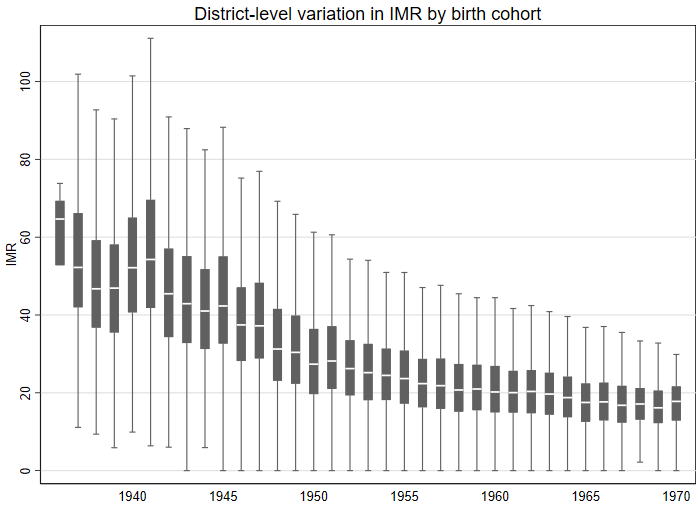}
  \caption{Trend and variation in infant mortality rates for England and Wales; 1935--1970}
  \label{fig:IMR}
\end{figure}

To better understand the temporal and spatial variation in infant mortality rates, \autoref{fig:IMR} presents a box plot of the district-level variation in infant mortality rates over time. The white line in the centre of the box is the median, with the box representing the inter-quartile range. This shows two main points. First, there has been a large reduction in the infant mortality rate over time, from approximately 50--60 infant deaths per 1000 live births in the late 1930s to just under 20 infant deaths per 1000 live births in the 1960s. Second, although the inter-quartile range reduces over time, it remains relatively large, even in the late 1960s, suggesting there is sufficient variation in infant mortality rates across districts to investigate its effect on later life outcomes.

To explore the extent to which the infant mortality rate is correlated with other socio-economic indicators, we merge district-level data on social class from the UK Censuses in 1951, 1961 and 1971 to the district-level data on infant mortality rates in these years from the Registrar General reports. Social class is measured as the proportion of households in each district that is of social class I (professional), II (managerial/technical), III (skilled non-manual/manual), IV (semi-skilled manual) and V (unskilled manual). We also link data on illegitimacy; recently proposed as a proxy for social class \citep{Luehrmann2018}. We then regress the infant mortality rate on the illegitimacy rate (i.e., the number of illegitimate children per 1000 live births) and the social class indicators. \autoref{tab:IMR_SES} presents the estimates, showing a strong social gradient in infant mortality, with districts that have larger proportions of individuals in lower social classes experiencing significantly higher infant mortality rates. This is true for the pooled sample (column 1), as well as for the individual Census years (columns 2--4), though the magnitude reduces over time and it is no longer significant in 1971. The strong correlation between the infant mortality rate and social class is important, as this suggests that the former is likely to partially capture different socio-economic indicators of the district. We find some suggestive evidence that the illegitimacy rate is positively correlated with infant mortality, though only in 1971. 

\bigskip
\begin{table}[ht]
\caption{Correlations between district level infant mortality rates, social class and illegitimacy}
\centering
{\scriptsize
\begin{tabular}{lcccccccccccccccccccc}
\toprule
\input{tab/IMR_SES}
\bottomrule
\addlinespace[.75ex]
\end{tabular}
}
\label{tab:IMR_SES}
\caption*{\noindent\scriptsize Notes: Column (1) additionally includes census year dummies. Social class III is the reference category. Robust standard errors, clustered by district, in parentheses. * $p < 0.1$, ** $p < 0.05$, *** $p < 0.01$.}
\end{table}

\subsection{The Polygenic Score}
\label{sec:descr_PGS}

\begin{figure}[ht]
  \centering
  \includegraphics[width=0.3\linewidth]{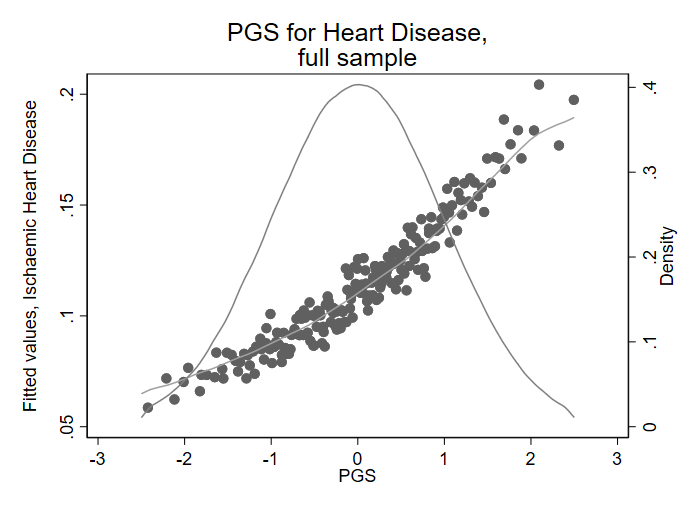}
  \includegraphics[width=0.3\linewidth]{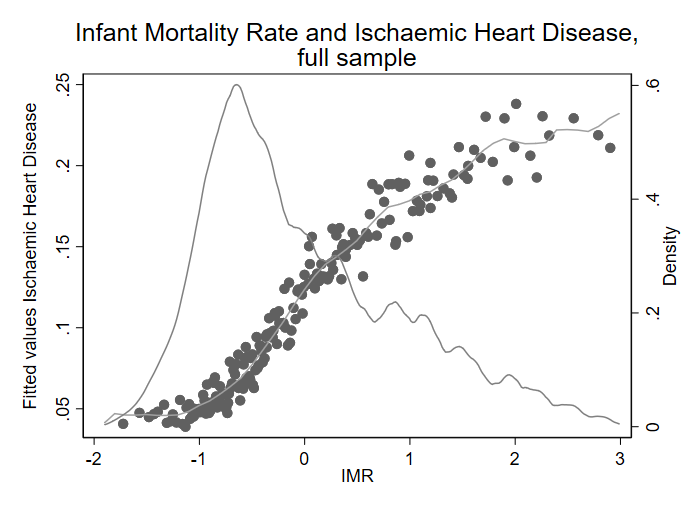} 
  \includegraphics[width=0.3\linewidth]{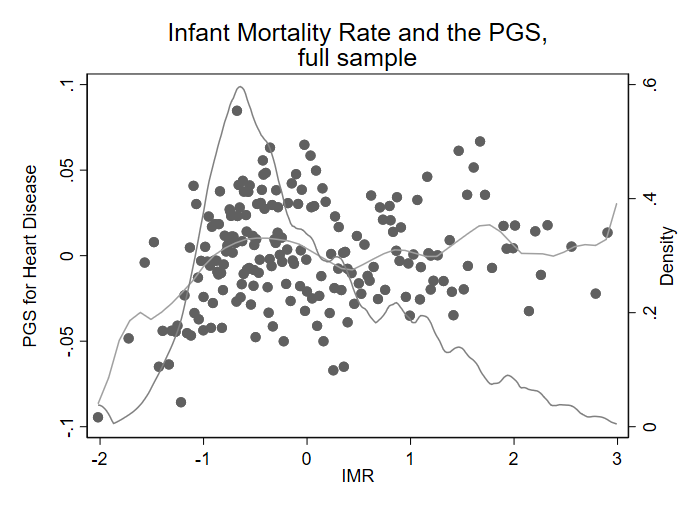} \\
  \includegraphics[width=0.3\linewidth]{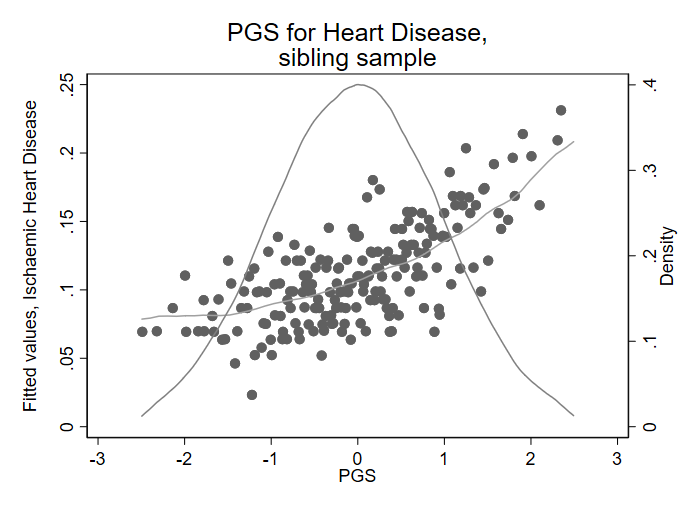}
  \includegraphics[width=0.3\linewidth]{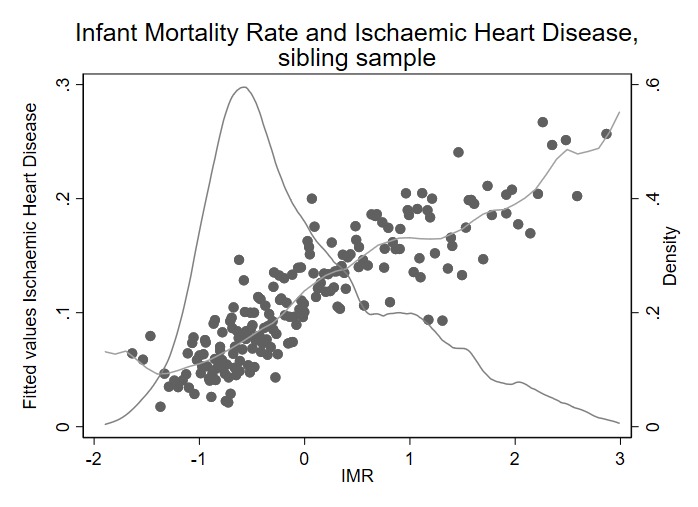} 
  \includegraphics[width=0.3\linewidth]{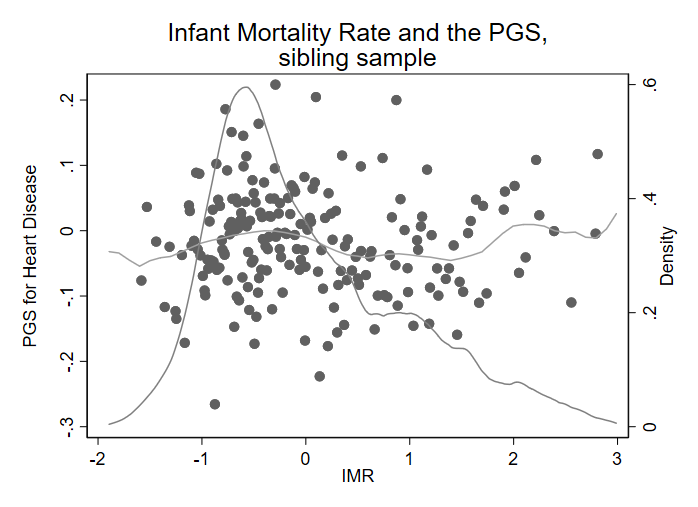} \\
  \includegraphics[width=0.3\linewidth]{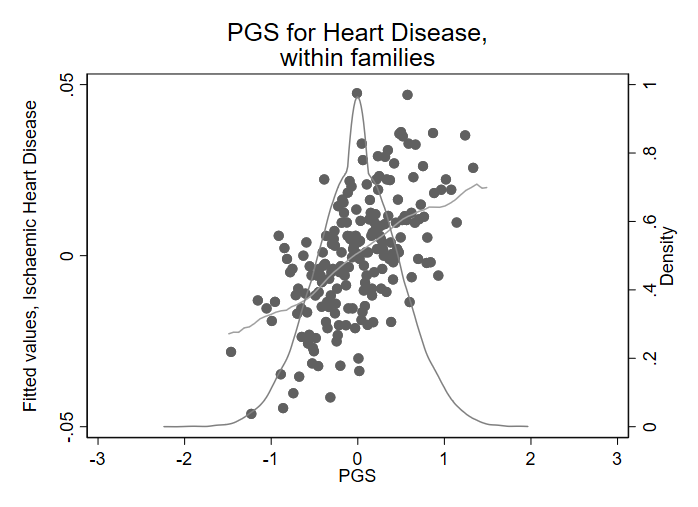}
  \includegraphics[width=0.3\linewidth]{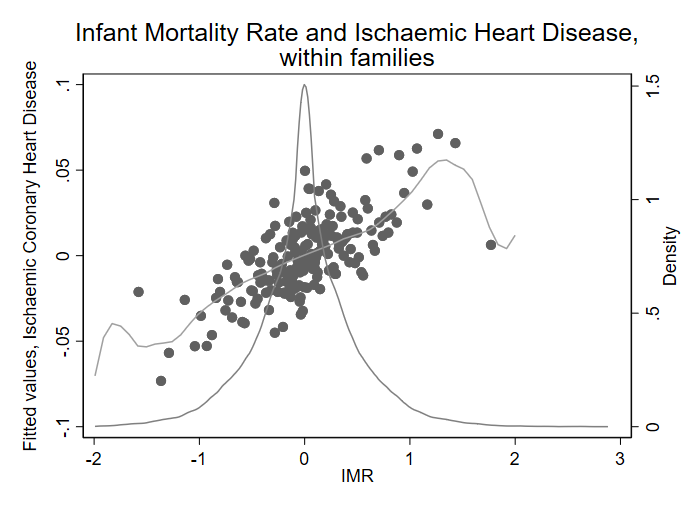} 
  \includegraphics[width=0.3\linewidth]{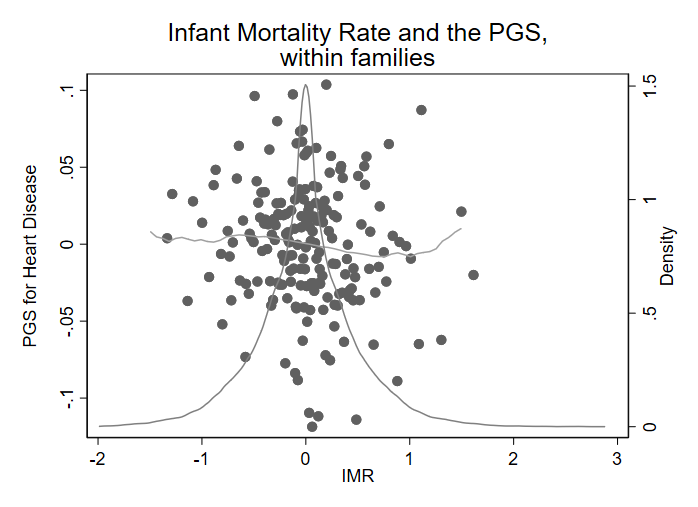} \\
  \caption{Correlation between the polygenic score, infant mortality rates, and ischaemic heart disease in the full estimation sample (row 1), the sibling sample (between families; row 2) and sibling sample (within families, row 3).}
\label{fig:DensityPGS}
\end{figure}

We next provide descriptive statistics on the polygenic score and how it correlates with the outcome of interest. The top left hand graph of \autoref{fig:DensityPGS} presents the density of the (standardized) polygenic score for heart disease. We divide the PGS into 200 bins, and plot the average outcome (i.e., the probability of being diagnosed with ischaemic heart disease) for each bin. These are the black dots. The line through the dots is obtained from a kernel-weighted local polynomial regression of the outcome on its polygenic score. This shows a strong correlation between the polygenic score and ischaemic heart disease. We report the statistical strength of this relationship in regressions below. The figures also show some suggestion of convexity in the relationship between the polygenic score and the outcome. We model this directly below and explore the sensitivity of our findings to linear specifications of the polygenic score in the robustness analysis.

The middle graph in the top row of \autoref{fig:DensityPGS} shows a similar plot, but with the Infant Mortality Rate (IMR) on the horizontal axis. This shows a strong correlation between the infant mortality rate at birth and the likelihood of being diagnosed with ischaemic heart disease in later life. However, the graph suggests that the relationship is somewhat non-linear, with little difference in ischaemic heart disease for those born in areas characterised by low (i.e. more than 1 standard deviation below the mean) rates of infant mortality, but with increasing risk of ischaemic heart disease among those with (standardised) infant mortality rates above -1.

The top right hand graph in \autoref{fig:DensityPGS} shows the extent to which the PGS and the infant mortality rate are correlated with one another. Positive (negative) gene-environment correlation -- $rGE$ -- indicates that individuals with a higher genetic predisposition are more (less) likely to be observed in environments with higher infant mortality rates. The figure shows some suggesting evidence of positive $rGE$, with districts exposed to higher than average infant mortality rates also having higher than average polygenic scores, although -- as we show below -- this correlation does not survive in our sibling sample. 

The middle set of three graphs in \autoref{fig:DensityPGS} show the same relationships as above, but on the sibling sample only. This still shows a positive correlation between the polygenic score and ischaemic heart disease (figure on the left). As we show in the regressions below, the strength of this relationship is similar in this reduced sample. The correlation between the infant mortality rate and ischaemic heart disease (graph in the middle) is again positive, but does not suggest strong non-linearities. Furthermore, we find no evidence of $rGE$ in the reduced sibling sample (graph on the right), with the slope of the line being close to zero.

Finally, the bottom three graphs only exploit \textit{within-family} variation. In other words, we use the UK Biobank sibling sample and correlate the \textit{residuals} of the variables shown on the axes, after taking out the family fixed effects. This still shows a positive correlation between the polygenic score and ischaemic heart disease (bottom left figure), though -- as we show below -- the magnitude of the effect is slightly smaller, since the within-family analysis accounts for shared environments and parental genotypes \citep[see e.g.][]{Selzam2019,koellinger2018using,kong2018nature,LeeEtAl2018}. 
Accounting for time-invariant characteristics within families, we still find a positive correlation between the infant mortality rate and ischaemic heart disease, as shown in the bottom middle graph, though this is effects is again attenuated. Finally, we find no evidence of gene-environment correlation ($rGE$; bottom right graph). This is reassuring, as it suggests we are identifying true gene-environment interactions, rather than spurious gene-gene or environment-environment interactions (for more detail on the interpretation of the $G \times E$ coefficient in the presence of $rGE$, see \cite{Biroli2022}).


\section{Empirical strategy}\label{sec:methods}
\paragraph{Baseline specification:}
To explore the relationship between ischaemic heart disease, the infant mortality rate at birth, the polygenic score and their interaction, we write the baseline empirical specification as:
\begin{align}
\label{eq:Eq}
Y_{id} = \alpha & + \beta_1 IMR_{d,t=0} + \beta_2 IMR^2_{d,t=0} + \beta_3 PGS_i + \beta_4 PGS^2_i \nonumber \\
& + \beta_5 PGS_i \times IMR_{d,t=0} + \beta_6 PGS^2_i \times IMR^2_{d,t=0} \nonumber \\
& + \gamma \mathbf{X_i} + f_1(IMR,\mathbf{X}) + f_2(PGS,\mathbf{X}) + u_{id}
\end{align}

where $Y_{id}$ is equal to one if individual $i$, born in district $d$, has been diagnosed with ischaemic heart disease, and zero otherwise. The district-level infant mortality rate in the year of birth (i.e., $t=0$) is given by $IMR_{d,t=0}$, which is standardised to have mean 0 and standard deviation 1. The relevant polygenic score is given by $PGS_i$, also standardised to have mean 0 and standard deviation 1. The gene-environment interaction is denoted by $PGS_i \times IMR_{d,t=0}$. We follow \citet{Biroli2022} and include non-linear terms of $PGS_i$, $IMR_{d,t=0}$, as well as $PGS_i \times IMR_{d,t=0}$, though we explore the robustness of our results to linear specifiations in \hyperref[sec:robustness]{Section~\ref*{sec:robustness}} below; we show that this does not change our main conclusions. 

Within $\mathbf{X_i}$, we control for the individual's gender and we include dummies for each year-month of birth (i.e., [(12 months $\times$ 38 years)-1] separate dummy variables). The latter will account for the fact that older individuals are more likely to be exposed to higher infant mortality rates, and are also more likely to have worse health, on average. Hence, including dummies for each year-month of birth is the most flexible specification to account for individuals' age at the \textit{monthly}-level. We also include the first 10 principal components of the genetic relatedness matrix to control for any remaining genetic differences across ancestry groups, as is custom in the literature \citep[see e.g.][]{price2006principal}.\footnote{Although it is not necessary to control for the principal components in the within-family analysis, we include them to allow the comparison to the between-family analysis. Excluding them does not affect our conclusions.} Finally, the function $f_1(IMR,\mathbf{X})$ and $f_2(PGS,\mathbf{X})$ denotes interactions between $IMR_{d,t=0}$ and $\mathbf{X_i}$ and between $PGS_i$ and $\mathbf{X_i}$ respectively \citep[as in][]{Keller2014}. The error is denoted by $u_{id}$; we report heteroskedasticity-robust standard errors, clustered by either by district (in full sample) or by family and district (in the family fixed effects analysis below).

We start by estimating the relationship between the local infant mortality rate in the year of birth and later life ischaemic heart disease (i.e. Equation (\ref{eq:Eq}) without $PGS_i$, $PGS^2_i$ and the interaction terms). This analysis is meant as a replication of the original Barker hypothesis, showing the strength of the relationship between infant mortality rates in the year and region of birth and ischaemic heart disease, and allowing for a non-linear (quadratic) effect. However, note that, in contrast to \cite{BarkerOsmond1986}, our outcome of interest is measured at the \textit{individual} level, as opposed to the \textit{regional} level. We next explore the statistical relationship between the outcome of interest and the relevant polygenic score (i.e., Equation (\ref{eq:Eq}) without $IMR_{d,t=0}$, $IMR^2_{d,t=0}$ and the interaction terms), aiming to corroborate the results from existing GWAS that genetic scores are predictive of heart disease, as well as to quantify this relationship. 

Finally, by including the main effects as well as their interactions, as in Equation (\ref{eq:Eq}), we explore whether being born in a district characterised by low infant mortality rates can reduce not only the \textit{mean} disease prevalence in older age, but also its \textit{variation} predicted by the polygenic score. Or vice versa, whether an individual's polygenic score exacerbates or protects against the effects of being born in high infant mortality rate districts. 

Since we are exploring the effects of early life circumstances (and in particular: mortality) on later life health, it is important to take into account the potential for selection. Indeed, a high infant mortality rate implies that only a selected sample remains (i.e., those who survived). Assuming that the survivors are healthier than those who did not survive, we are likely to underestimate the effects of interest. 

\paragraph{District Fixed Effects:} 
Equation (\ref{eq:Eq}) compares individuals born in districts with high infant mortality rates to those born in districts with lower infant mortality rates, conditional on the other covariates including the year-month of birth. However, districts with a high infant mortality rate may be systematically different from districts with a low infant mortality rate. Indeed, \autoref{tab:IMR_SES} shows that the population in high infant mortality rate districts is likely to be of lower socio-economic class. To take this into account, we next add district fixed effects, only exploiting variation in infant mortality rates \textit{within} districts over time. Such district fixed effects specifications ensure that the parameters are identified only off of \textit{within}-district variation, comparing individuals born in the same district, yet exposed to different infant mortality rates depending on their year of birth. 

\paragraph{Family Fixed Effects:}
Whereas district fixed effects go a long way in controlling for socio-economic differences across districts, the socio-economic composition of districts may change over time. As such, an increase in the infant mortality rate may still reflect an increase in regional poverty, which may be driving the results from the regressions above. We therefore also exploit the fact that the UK Biobank includes a sample of full siblings, meaning we can look at variation within full-sibling pairs from the same family. The family fixed effects specification is, given by:
\begin{align}
\label{eq:Eq_FFE}
Y_{ijd} = \alpha & + \beta_1 IMR_{ijd,t=0} + \beta_2 IMR^2_{ijd,t=0} + \beta_3 PGS_{ij} + \beta_4 PGS^2_{ij} \nonumber \\
& + \beta_5 PGS_{ij} \times IMR_{ijd,t=0} + \beta_6 PGS^2_{ij} \times IMR^2_{ijd,t=0} \nonumber \\
& + \gamma \mathbf{X_{ij}} + f(IMR,PGS,\mathbf{X}) +\eta_j + u_{ijd}
\end{align}

where $\eta_j$ are the family fixed effects; the other variables are defined above. As such, Equation (\ref{eq:Eq_FFE}) exploits the fact that some individuals are born in years with low infant mortality rates, but their siblings, who largely share the same family environment, may be born in years with higher or lower infant mortality rates. Hence, this specification exploits variation in infant mortality, holding any other time-invariant (observed or unobserved) family characteristics fixed and as such accounts for any confounders that differ \textit{between} households that may bias the estimates from Equation (\ref{eq:Eq}).

An additional advantage of the sibling sample is that estimation of the genetic effects is purged from endogeneity concerns. That is, even though one's genotype is fixed at conception and therefore by definition not affected by early-life circumstances, parental genotypes can act as potential confounders. More specifically, the parental genotype can shape the environment in which children are raised, also known as `\textit{genetic nurture}', and therefore act as a potential confounder in the relationship between one's own genotype and the outcome \citep[see, e.g.,][]{kong2018nature,wertz2019genetics}. Whilst this may be more of a concern for polygenic scores related to socio-economic outcomes such as educational attainment, we cannot rule out endogeneity of the polygenic score for heart disease. Controlling for family fixed effects exploits only variation in genetic variants \textit{within} sibling pairs, the allocation of which is randomly assigned by Mendel's law. 

In sum, given the random inheritance of genetic variants within families, the coefficients $\beta_3$ and $\beta_4$ in Equation (\ref{eq:Eq_FFE}) capture the direct genetic effect. Moreover, since our environmental measure -- the infant mortality rate in the district and year of birth -- is uncorrelated with our polygenic score in the within-family analysis (see Section \ref{sec:descr_PGS}), we are able to estimate a genuine `$G \times E$' interaction, as opposed to spurious `$G \times G$' or `$E \times E$' due to e.g., gene-environment correlations or genetic nurture.\footnote{We show in \autoref{fig:RobustnessPGS} in Appendix A that the infant mortality rate is uncorrelated to a wide range of \textit{other polygenic scores}, and similarly, that the polygenic score for heart disease is uncorrelated with a range of \textit{alternative early life environments} (\autoref{fig:RobustnessE}). Indeed, finding evidence of systematic $rGE$ between the infant mortality rate and alternative polygenic scores would suggest that our $G \times E$ estimate may in fact be picking up a gene-\textit{gene} interaction effect. Vice versa, strong correlations between the polygenic score for heart disease and alternative early life environments would suggest that our $G \times E$ estimate may instead be capturing an \textit{environment}-environment interaction effect. Our analysis shows no strong evidence of $rGE$ in either case, with a wide range of polygenic scores and early life environments, reinforcing the argument that we are identifying genuine $G \times E$ interactions.}

Having said that, we acknowledge that we cannot claim to estimate the \textit{causal} effect of the infant mortality rate in one's district and year of birth. Even though the district-- and family fixed effects specifications allow us to rule out that the infant mortality rate is merely picking up district- or family-specific time-invariant (socioeconomic) conditions, and so we are confident that the infant mortality rate reflects early-life environmental conditions, we cannot open the black box of \textit{which} early-life environmental factors are the driving causal mechanism. In other words, it is likely that the infant mortality rate proxies for other early life `environments'.

Finally, whereas the sibling sample is helpful to reduce endogeneity concerns, a downside is that it is significantly smaller compared to the main sample. There are approximately 33,000 siblings in our analysis sample; substantially less than the full sample. This has two (related) implications, both leading to a loss of power. First, using the smaller sample will directly inflate the standard errors. Second, including family fixed effects means we are only exploiting variation \textit{within} families. As most siblings are born relatively close together and infant mortality rates do not change dramatically over the course of a few years within a given district, there is relatively little variation in infant mortality rates over time within the same family. Hence, for both these reasons, the within-family analysis has much less power than the between-family analysis, and this will be reflected in the standard errors. We therefore focus more on the magnitude of the estimates, as opposed to its precision.


\section{Results}
\label{sec:results}

\paragraph{Baseline specification:} Column 1 in \autoref{tab:IHD} presents the estimates from a simple regression of the binary indicator of ischaemic heart disease on the infant mortality rate in the district and year of birth as well as its square, controlling for gender and year $\times$ month of birth dummies. This confirms the results in \cite{BarkerOsmond1986}, showing a significant relationship between adverse early life conditions and later life cardiovascular health. In fact, the estimates suggest an inverse U-shaped association. Column 2 shows the predictive power of the polygenic score, corroborating the findings from existing GWAS and showing a significant positive relationship between the polygenic score for heart disease and the individual diagnosis, which is stronger for those with higher polygenic scores. Including the infant mortality rate and the polygenic score simultaneously, as in column 3 and column 4, shows effects of similar magnitude. Column 4 further adds the interaction term between genetic variation and the early life environment, showing a positive and significant effect, with no evidence of non-linearities in the squared interaction. In other words, an increase in the infant mortality rate in the district and year of birth increases the probability of being diagnosed with ischaemic heart disease by more for those with a high polygenic score for heart disease.

\bigskip
\begin{table}[ht]
\caption{Gene-environment interplay for ischaemic heart disease}
\centering
{\scriptsize
\begin{tabular}{lcccccccccccccccccccc}
\toprule
\input{tab/IHD.tex}
\bottomrule
\addlinespace[.75ex]
\end{tabular}
}
\label{tab:IHD}
\caption*{\noindent\scriptsize Notes: Covariates include gender and all year$\times$month of birth dummies. `Mean' is the mean of the dependent variable. Robust standard errors clustered by district in parentheses. * $p < 0.1$, ** $p < 0.05$, *** $p < 0.01$.}
\end{table}

\begin{figure}[ht]
  \centering
  \includegraphics[width=0.7\linewidth]{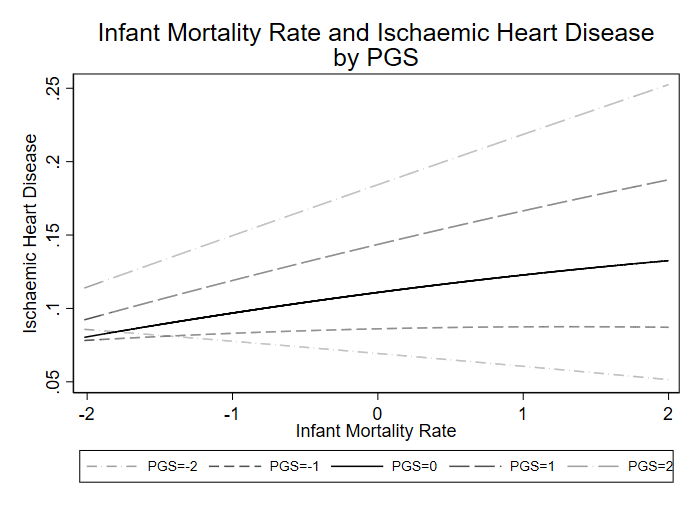}
  \caption{$G \times E$ interplay for Ischaemic Heart Disease; full sample}
  \label{fig:IHDslopes}
\end{figure}

\autoref{fig:IHDslopes} graphically presents the regression results from column 4 in \autoref{tab:IHD}, showing a fanning out of the regression lines. In other words, a less advantageous early life environment is associated with an increased probability of being diagnosed with ischaemic heart disease. However, the magnitude of this effect increases substantially with individuals' polygenic score: the association with the infant mortality rate is much stronger for those with higher genetic risk for developing ischaemic heart disease. The graph also suggests that genetic risk plays a minor role in the probability of being diagnosed with ischaemic heart disease among those exposed to low infant mortality rates at birth, with little variation in the outcome of interest across polygenic scores for those born into low infant mortality rate districts. This suggests that one's genetic predisposition matters less for the development of ischaemic heart disease among those born in advantageous, healthier environments.

\begin{figure}[ht]
  \centering
  \includegraphics[width=0.45\linewidth]{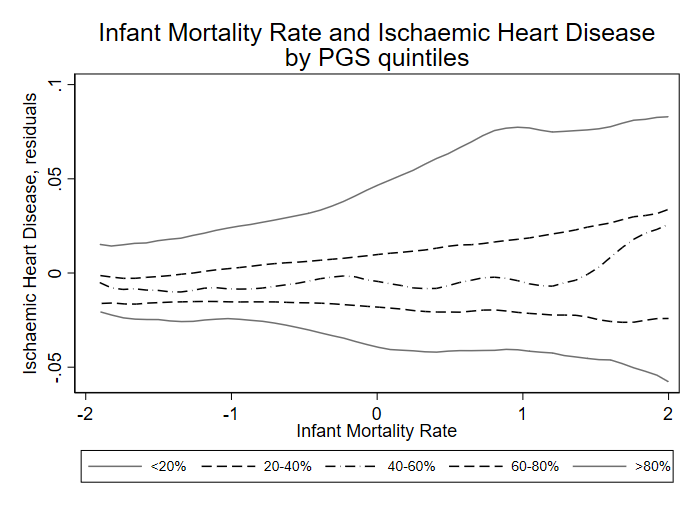}
  \includegraphics[width=0.45\linewidth]{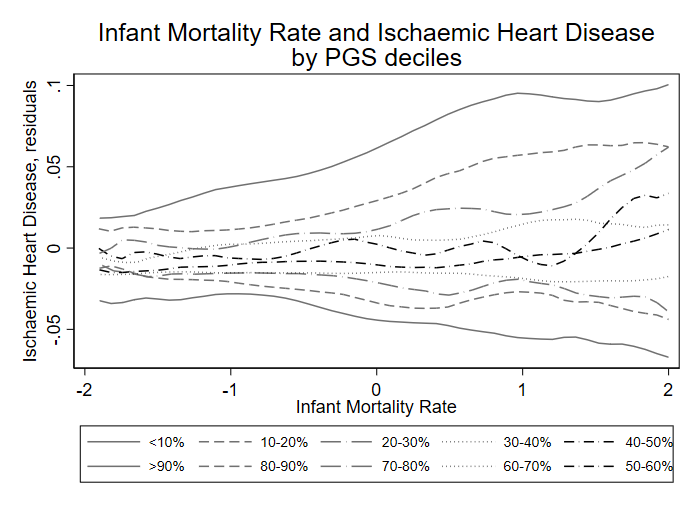}
  \caption{$G \times E$ interplay for Ischaemic Heart Disease, local polynomial plots; full sample}
  \label{fig:IHDslopes_groups}
\end{figure}

To explore more flexible non-linearities in the gene-environment interaction effect, we next plot the relationship between the \textit{residualised} outcome and the infant mortality rate at birth using local polynomial plots, where the residual is obtained from a regression of ischaemic heart disease on the covariates listed in Equation (\ref{eq:Eq}). The left hand panel of \autoref{fig:IHDslopes_groups} shows this for the quintiles of the polygenic score, whereas the panel on the right presents it for the polygenic score deciles. This shows similar trends to those in \autoref{fig:IHDslopes}, illustrating a diverging pattern as the infant mortality rate increases. Again, the figure suggests that one's genetic predisposition makes little difference in the probability of developing ischaemic heart disease among those born in more advantageous environments, yet plays a major role when exposed to an adverse early life environment. 

\paragraph{District Fixed Effects:} To explore the extent to which these coefficients are capturing unobserved differences across districts in infant mortality rates and later life ischaemic heart disease, column 5 in \autoref{tab:IHD} adds district of birth fixed effects, taking into account that, e.g., districts with high infant mortality rates are systematically poorer than districts with lower infant mortality rates. Hence, we now only exploit variation in infant mortality rates \textit{within} districts over time, comparing individuals born in the same district but exposed to different rates of infant mortality due to their year of birth. The findings suggest that over half of the main infant mortality rate effect is due to time-invariant differences between districts. A one standard deviation increase in infant mortality rates in the district and year of birth is associated with a 0.5 percentage point (4\%) increase in the probability of being diagnosed with ischaemic heart disease (with its square being insignificantly different from zero). The coefficients on the polygenic score and the interaction do not change with the addition of district fixed effects, suggesting there is no systematic variation in these variables across districts, once we account for the main effect of the infant mortality rate.

\paragraph{Family Fixed Effects:} \autoref{tab:IHD_Sibs} reproduces the regression results for the full sample without district fixed effects (columns 1--3); the results for the reduced sibling sample without family fixed effects (columns 4--6), and the results for sibling sample including family fixed effects (columns 7--9). Reducing the sample size to include only siblings causes a substantial increase in all standard errors, yet does not lead to big changes in the estimated coefficients of the infant mortality rate and the polygenic score. However, the estimate on the interaction halves, suggesting that the sibling sample is different from the full UK Biobank. Indeed, the sibling sample is more likely to be born in urban areas, with more limited representation of the rural districts in England and Wales. Nevertheless, it remains statistically significant, even on the reduced sample. 

Moving to the estimates from the family fixed effects specifications (columns 7--9), we find a very small reduction in the parameter estimate on the polygenic score, suggesting that demography and indirect genetic effects do not play a large role for ischaemic heart disease. However, the coefficient on the infant mortality rate approaches zero and is no longer significant. This therefore suggests that the infant mortality rate partially proxies for other unobserved characteristics that vary \textit{between} families. Once these are accounted for via the family fixed effects, the coefficient on the infant mortality rate is no longer significant. The interaction effects, however, remain very similar to the point estimate in the between-family analysis (column 6). In other words, the results suggest that the infant mortality rate does not increase the probability of being diagnosed with ischaemic heart disease for those with an average polygenic score, yet that it does increase this probability for those with a high polygenic score. In the other direction, a one standard deviation increase in the polygenic score increases the probability of being diagnosed with ischaemic heart disease, and this effect is larger for those exposed to higher infant mortality rates at birth.

\bigskip
\begin{table}[ht]
\caption{Restricting the sample size and including family fixed effects}
\centering
{\scriptsize
\begin{tabular}{lcccccccccccccccccccc}
\toprule
\input{tab/IHD_Sibs_noDFE.tex}
\bottomrule
\addlinespace[.75ex]
\end{tabular}
}
\label{tab:IHD_Sibs}
\caption*{\noindent\scriptsize Notes: Columns (1)-(3) show robust standard errors clustered by district on the full sample. Columns (4)--(6) show robust standard errors clustered by district on the sibling sample. Columns (7)--(9) use two-way clustering by family and district on the sibling sample. * $p < 0.1$, ** $p < 0.05$, *** $p < 0.01$.}
\end{table}

\begin{figure}[ht]
  \centering
  \includegraphics[width=0.45\linewidth]{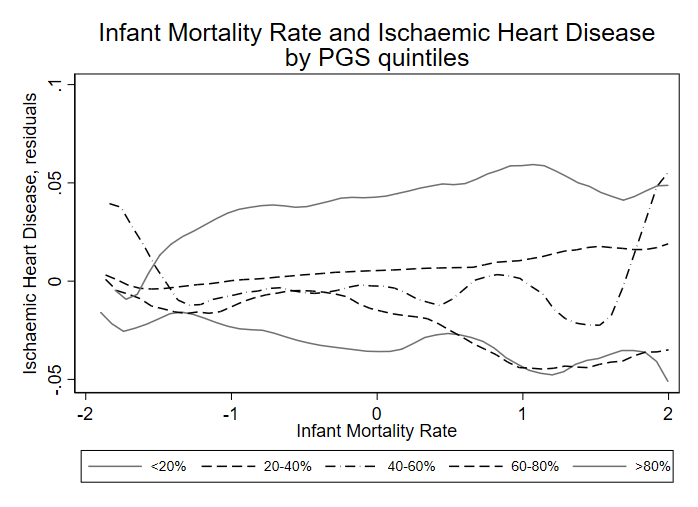}
  \includegraphics[width=0.45\linewidth]{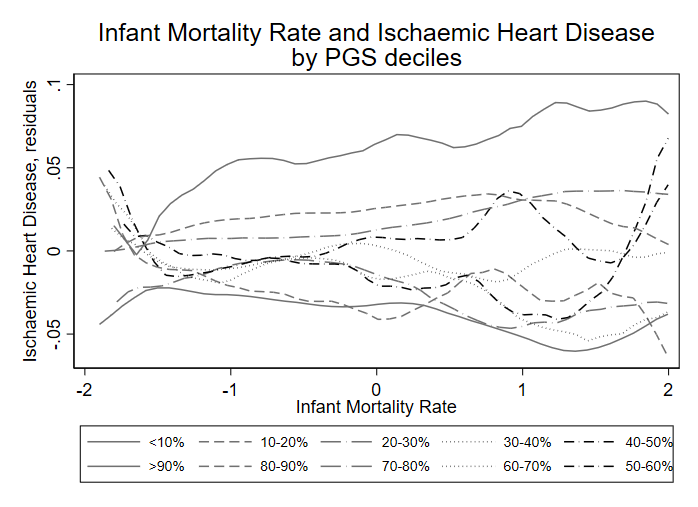}
  \caption{$G \times E$ interplay for Ischaemic Heart Disease, local polynomial plots; sibling sample, accounting for family fixed effects.}
  \label{fig:IHDslopes_groups_FFE}
\end{figure}

\autoref{fig:IHDslopes_groups_FFE} shows the non-linear local polynomial plots from the residualised outcome, as in \autoref{fig:IHDslopes_groups} above, but now also residualised for family fixed effects. This again shows rising divergence in ischaemic heart disease with increasing infant mortality rates. The probability of developing ischaemic heart disease is very similar for those exposed to low infant mortality rates at birth, with one's polygenic score being more important for those exposed to high infant mortality rates.


\paragraph{Potential mechanisms:} 
To examine potential mechanisms for the significant \textit{interaction} between the polygenic score for heart disease and the infant mortality rate in early life, we investigate whether and how the main and interaction effects are associated with specific behaviours and other outcomes that also associate with heart disease, including BMI, blood pressure, height, drinking and smoking. Indeed, if the main and interaction effects show no association with these alternative outcomes, they cannot be a mediator in the relationship between infant mortality, polygenic scores and heart disease. We define smoking and drinking as dummy variables that equal to one if the individual currently smokes or drinks, or has done so in the past, and zero otherwise.

\autoref{tab:mediators} presents the estimates, where the first column for each outcome only controls for the covariates specified in Equation (\ref{eq:Eq}) and the second additionally accounts for family fixed effects, as in Equation (\ref{eq:Eq_FFE}). All analyses are based on the sibling sample and specify the polygenic score for heart disease as the variable capturing the genetic component. Examining first the estimates that do not control for family fixed effects, we find that the infant mortality in one's year and district of birth has an inverse U-shaped association with BMI and systolic blood pressure, as well as a U-shaped relationship with height in adulthood. There are no strong differences in drinking and smoking among those born in areas characterised by different infant mortality rates. Once we control for family fixed effects, however, these associations reduce in magnitude and are generally no longer significantly different from zero, though there is some evidence of marginal non-linearities in BMI and systolic blood pressure. 

Furthermore, we find that the polygenic score for heart disease predicts variation in BMI, blood pressure, height, alcohol and smoking status, with the estimates suggesting these associations are generally linear. When controlling for family fixed effects, the estimates reduce, but they all remain significantly different from zero. Finally, the coefficients on the $G \times E$ terms in the analyses that do not include family fixed effects are significant only for BMI and diastolic blood pressure, with no evidence of any non-linearities, where a higher polygenic score for heart disease is protective against the adverse effects of increased infant mortality rates. 
However, including family fixed effects renders them insignificantly different from zero. In other words, we find no strong evidence that BMI, blood pressure, height, drinking, and smoking mediate the relationship between polygenic scores and infant mortality rates in the development of ischaemic heart disease.

\begin{landscape}
\begin{table}[ht]
\caption{Potential mediators in the gene-environment interplay for ischaemic heart disease}
\centering
{\scriptsize
\begin{tabular}{lcccccccccccccccccccc}
\toprule
\input{tab/mediators.tex}
\bottomrule
\addlinespace[.75ex]
\end{tabular}
}
\label{tab:mediators}
\caption*{\noindent\scriptsize Notes: BMI is measured as weight (in kilograms) divided by height (in metres squared). DBP add SBP denote diastolic and systolic blood pressure, respectively, and are measured in mmHg (millimeters of mercury). Height is measured in cm. Alcohol and smoking are dummy variables indicating whether the individual currently smokes or drinks. `Mean' is the mean of the dependent variable. Robust standard errors clustered by district (column 1) and family and district (column 2) in parentheses. * $p < 0.1$, ** $p < 0.05$, *** $p < 0.01$.}
\end{table}
\end{landscape}

\section{Robustness checks}
\label{sec:robustness}
This section explores the sensitivity of our main findings to a set of robustness checks. 
First, we explore whether there are gender differences in the relationship between the early life infant mortality rate and later life heart disease. 
Second, we test the sensitivity of our findings to controlling for additional covariates at the district level. 
Third, we explore the use of an alternative polygenic score for heart disease, constructed from the summary statistics of our tailor-made GWAS on the non-siblings (and non-relatives) of the UK Biobank. 
Fourth, we examine the robustness of our results to specifying the main and interaction effects as linear predictors (i.e., not allowing for non-linearities in $PGS$ and $IMR$).
Fifth, we compare our analysis to those that use the infant mortality rate in the years around birth (as opposed to the year \textit{of} birth) as the main variable of interest. 
Other than our first robustness check, all are based on the family fixed effects specification.

\subsection{Gender differences}
We start by exploring whether the relationship between ischaemic heart disease and infant mortality rates at birth is similar for men and women. \autoref{tab:IHDgender} shows slightly larger estimates for men compared to women, possibly reflecting a higher incidence of heart disease among men. In addition, we find that the polygenic score is more predictive among men, with the magnitude of the interaction term also being slightly larger. Nevertheless, for both groups do we find evidence of main as well as interaction effects for ischaemic heart disease. Furthermore, we find evidence of non-linearities of the interaction term for women, though not for men. 

\begin{table}[ht]
\caption{$G \times E$ interplay for ischaemic heart disease, by gender}
\centering
{\tiny
\begin{tabular}{lcccccccccc}
\toprule
\input{tab/IHDgender.tex}
\bottomrule
\addlinespace[.75ex]
\end{tabular}
}
\label{tab:IHDgender}
\caption*{\noindent\scriptsize Notes: These analysis are based on the full UK Biobank analysis sample, controlling for district fixed effects in columns (5) and (10). `Mean' is the mean of the dependent variable. Robust standard errors clustered by district in parentheses. * $p < 0.1$, ** $p < 0.05$, *** $p < 0.01$.}
\end{table}

\subsection{Controlling for additional district-level covariates}
As shown in \autoref{tab:IMR_SES}, the infant mortality rate is likely to capture variation in socio-economic characteristics. Although the district fixed effects capture some of this variation, it may not be able to account for time-varying characteristics that correlate with the infant mortality rate. We therefore next explore whether our analysis is robust to the inclusion of a set of variables that vary over time \textit{within} districts. For this, we include district-level illegitimacy, birth and death rates in the regressions, as measured during individuals' year of birth, aiming to capture some socio-economic differences between districts. Column (1) of \autoref{tab:IHDrobustness} replicates the main estimates from column (9) of \autoref{tab:IHD_Sibs} for comparison. Column (2) reports the estimates that control for the additional covariates, showing very similar coefficients on the infant mortality rate, the polygenic score, and the interactions, with no significant effects of the additional covariates.

\bigskip
\begin{table}[ht]
\caption{Robustness analyses of gene-by-environment interplay for ischaemic heart disease}
\centering
{\scriptsize
\begin{tabular}{lcccccccccccccccccccc}
\toprule
\input{tab/IHDrobustness.tex}
\bottomrule
\addlinespace[.75ex]
\end{tabular}
}
\label{tab:IHDrobustness}
\caption*{\noindent\scriptsize Notes: All estimates are from family-fixed effects specifications. `Covariates' refers to gender and year-month dummies; `PCs' are the first 10 principal components of the genetic relatedness matrix. `Additional controls' refer to the district-level covariates illegitimacy rate, birth rate and death rate, added in Column (2). Robust standard errors in parentheses, clustered by family and district throughout. Column (4) uses a polygenic score constructed from our own GWAS on non-siblings of the UK Biobank. All analyses control for the interactions between $IMR$ and $X$ and between $PGS$ and $X$. * $p < 0.1$, ** $p < 0.05$, *** $p < 0.01$.}
\end{table}

\subsection{Alternative polygenic score}
The polygenic score used in our main analysis is constructed from the summary statistics in \citet{nikpay2015comprehensive}. It is well-known, however, that the GWAS estimates are specific to the environmental context and demographic characteristics of the discovery sample \citep[see e.g.,][]{domingue2020interactions}. We therefore examine the sensitivity of our findings to the use of an alternative polygenic score, constructed using the summary statistics from our own tailor-made GWAS on the non-siblings of the UK Biobank sample (also excluding any other relatives of the sibling sample). The latter ensures that the discovery sample is independent of the analysis sample. It also ensures that the two samples are from the same environmental context, likely increasing the predictive power of the polygenic score. 

Column (3) of \autoref{tab:IHDrobustness} presents the results. Although the main effects of the infant mortality rate are not substantially affected, we find a somewhat larger main effect of the polygenic score. 
In addition, the estimate on the interaction term between the polygenic score and the infant mortality rate is larger. These findings confirm the existence of important interplay between the early life environment and individuals' genetic predisposition in the development of heart disease in older age.

\subsection{Modelling $G$ and $E$ as linear effects}
Our main analysis allows for non-linear (quadratic) effects in the infant mortality rate at birth, the polygenic score, as well as their interaction. To explore the robustness of our findings, we next re-estimate Equation (\ref{eq:Eq_FFE}) but restrict the main effects as well as the interaction terms to enter the function linearly. 
The estimates, presented in Column (4) of \autoref{tab:IHDrobustness}, are very similar to the specification that allows for non-linearities, and show evidence of an interaction effect between the infant mortality rate in one's year and district of birth and one's polygenic score in shaping individuals' probability of being diagnosed with heart disease in later life. 

\subsection{Exploring timing of the infant mortality rate}
\begin{figure}[ht]
  \centering
  \includegraphics[width=0.7\linewidth]{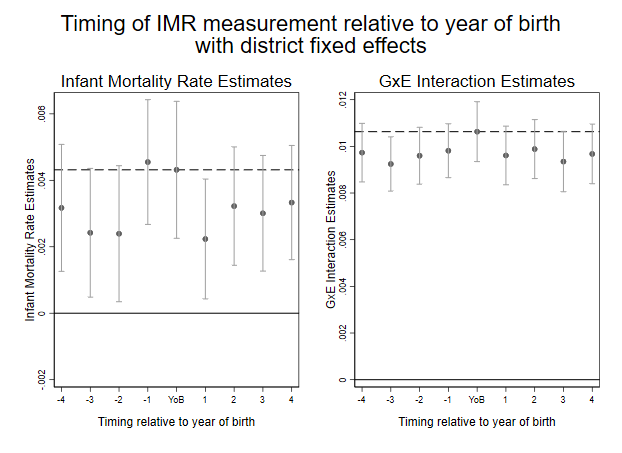}
  \caption{Exploring the timing of the infant mortality rate for Ischaemic Heart Disease; full sample, controlling for district fixed effects.}
  \label{fig:IHD_IMRtiming}
\end{figure}

\begin{figure}[ht]
  \centering
  \includegraphics[width=0.7\linewidth]{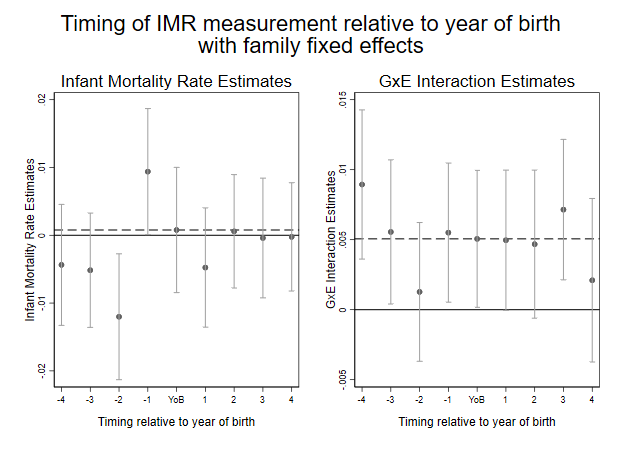}
  \caption{Exploring the timing of the infant mortality rate for Ischaemic Heart Disease; sibling sample, controlling for family fixed effects.}
  \label{fig:IHD_IMRtiming_FFE}
\end{figure}

\cite{BarkerOsmond1986} show similar correlations between ischaemic heart disease in adulthood and regional infant mortality rates, irrespective of whether neonatal or postneonatal infant mortality rates are used. We here explore whether our estimates are robust to different timings of the measurement of the infant mortality rate, ranging from four years before birth to four years after birth. \autoref{fig:IHD_IMRtiming} and \autoref{fig:IHD_IMRtiming_FFE} show the estimates, respectively, of the main infant mortality rate and the $G \times E$ interaction effects on the vertical axis, with the horizontal axis denoting the timing relative to the year of birth.\footnote{For ease of interpretation and plotting of the estimates, we here model the probability of being diagnosed with ischaemic heart disease as a function of $G$, $E$, and $G \times E$, but do not allow for the squared terms.} The estimates in \autoref{fig:IHD_IMRtiming} are based on a model that includes district fixed effects, whilst the estimates in \autoref{fig:IHD_IMRtiming_FFE} are based on a model that includes family fixed effects). The dashed horizontal line is the estimate from the analysis above, using the infant mortality rate measured in the year \textit{of} birth; the solid line is at zero.

The left panel of \autoref{fig:IHD_IMRtiming} suggests that it matters \textit{when} the infant mortality rate is measured. Using the infant mortality rate in the year of birth or the year prior to birth (we loosely refer to this as the year of pregnancy), the estimates are almost double the size compared to using earlier or later years,   
though we cannot statistically distinguish them from each other. The evidence is suggestive however that the prenatal and the first year of life are the most important in shaping later-life disease risk. The right panel of \autoref{fig:IHD_IMRtiming} shows that the interaction term is largest in magnitude for the infant mortality rate in the year of birth, although the other years show a remarkably consistent interaction effect that differs significantly from zero in all specifications.

\autoref{fig:IHD_IMRtiming_FFE} shows the estimates from the family fixed effects specification. These suggest no significant main effects of the infant mortality rate in the year of birth, but a significant positive effect for the infant mortality rate in the year of pregnancy. Taken at face value, this suggests that the prenatal period is more important than the neonatal period in shaping later-life disease risk. The interaction estimates, shown in the panel on the right, do not suggest that the timing of measurement matters, with the estimates being relatively constant across the different specifications. All show positive interaction effects, confirming that the early life environment interacts with individuals' genetic risk for heart disease. However, the fact that the estimates are similar across the different specifications suggests that the effects are not driven by factors that are specific to the year of birth. Instead, it looks like the infant mortality rate is a broader indicator of adverse environments in early life. 




\section{Conclusion}
\label{sec:concl}

The Barker hypothesis states that adverse circumstances in the prenatal and early life period can lead to disease in older age. We confirm this hypothesis in the UK Biobank, using a large sample of individuals in England and Wales born between the 1930s and 1970s. We then explore whether this relationship holds exploiting only variation within local areas or variation within families, and we investigate any heterogeneity of the association with respect to individuals' genetic predisposition. Our analysis shows a strong inverse U-shaped correlation between the local infant mortality rate at birth and later life heart disease. When using only variation in exposure to infant mortality rates and polygenic scores \textit{within siblings}, we show that the correlation is not significant for those with an \textit{average} polygenic score, but only exists among individuals with the highest genetic risk for developing heart disease. Although we still interpret our estimates as correlational rather than causal, our findings do show that gene-environment interplay in heart disease is strong and robust to the inclusion of district and family fixed effects.  

Our findings suggest that improvements in the early life environment reduce the role of genetic risk for heart disease. More specifically, being exposed to low infant mortality rates in one's year and region of birth reduces the risk of being diagnosed with ischaemic heart disease, with very little variation in this risk between those with high or low genetic predispositions for heart disease. In contrast, being exposed to high infant mortality rates increases the risk of ischaemic heart disease, and this risk is increased even further among those with a high genetic predisposition for heart disease. 

One important issue to take into account in the interpretation of the results is selection and scarring. First, it is well known that the UK Biobank is not representative of the UK population, with its participants being more likely to live in urban areas and being of higher socio-economic status, on average \citep{Fry2017}. This will affect the generalisability of our findings. Indeed, we cannot simply extrapolate our results to the full UK population, though our analysis shows that the positive correlation between infant mortality at birth and disease in old age also holds for this slightly higher SES (UK Biobank) population. If anything, we would expect this relationship to be stronger among individuals of lower SES, suggesting that we may underestimate the effects of interest. Second, individuals born in districts with high infant mortality rates may not be observed in the data, as they may have already passed away. If this is the case, our sample is a selected (healthier) sample of individuals, meaning that our estimates are likely to be a lower bound. Third, individuals born in districts with high infant mortality rates may have been `scarred' in early life because of their exposure to an adverse environment. In fact, this is one of the potential mechanisms suggested in the literature, that adverse conditions in early life may lead to permanent changes in the body structure, physiology and metabolism. Indeed, this is what we would likely be capturing in the analysis.

Our findings have at least two implications. First, our findings clearly show that `genetic determinism', or the idea that human behaviour is controlled by an individual's genes with no role for non-genetic/environmental influences, is incorrect. Indeed, we find strong evidence that `nature' and `nurture' complement each other. 
Our findings therefore suggest a more nuanced understanding of later life health, in which the effect of one's genetic predisposition depends on the environment in which individuals live and grow up \citep[see also, e.g.,][]{CaspiEtAl2010}. This suggests that improving the early life environment not only reduces the \textit{mean} prevalence in heart disease, but also the \textit{variation} in heart disease for those with different genetic predispositions. This is an interesting contrast with findings for education and cognition outcomes, where genetic effects are typically weaker in deprived environments \citep{heath1985education,tucker2016large,harden2021genetic}. 

Second, despite the fact that our study looks at infant mortality rates between the 1930s and 1970s, our findings do have policy relevance. Indeed, they suggest that improving the early life environment for children born now, will likely positively affect individuals' future health. Although we cannot pinpoint exactly what the infant mortality rate captures, and therefore what the main cause is of the change in disease prevalence in older age, our findings do suggest that a general improvement in such early life circumstances have long-run effects on the health and well-being of the population, in particular among those with high genetic risk for developing heart disease.


\clearpage

\footnotesize
\bibliographystyle{chicago}
\onehalfspacing
\bibliography{References}


\clearpage
\appendix 
\setcounter{figure}{0}
\setcounter{table}{0}
\renewcommand{\thetable}{A.\arabic{table}}
\renewcommand{\thefigure}{A.\arabic{figure}}
\section*{Appendix A: Gene-environment correlation}\label{sec:AppendixA}

\doublespacing 

\begin{figure}[ht]
  \centering
  \includegraphics[width=0.45\linewidth]{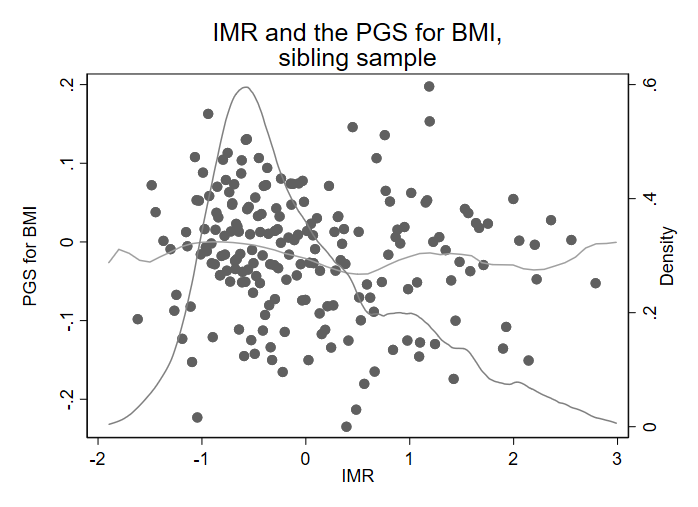}
  \includegraphics[width=0.45\linewidth]{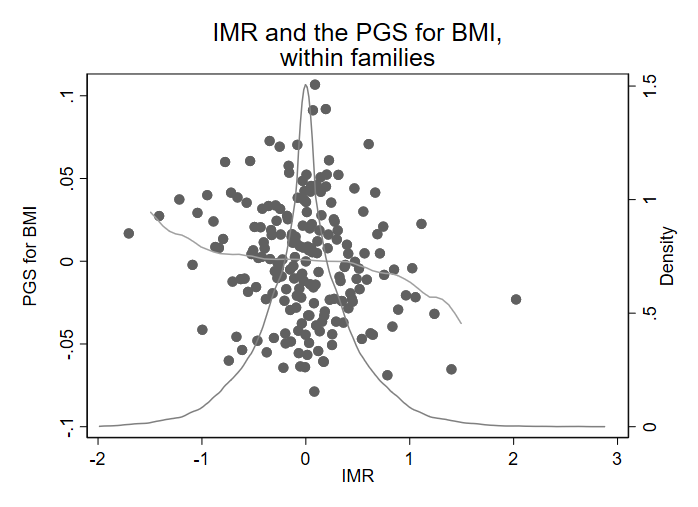} \\
  \includegraphics[width=0.45\linewidth]{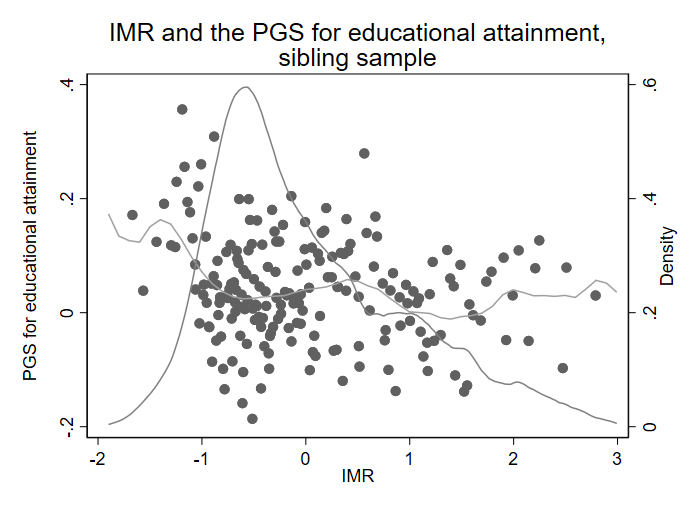} 
  \includegraphics[width=0.45\linewidth]{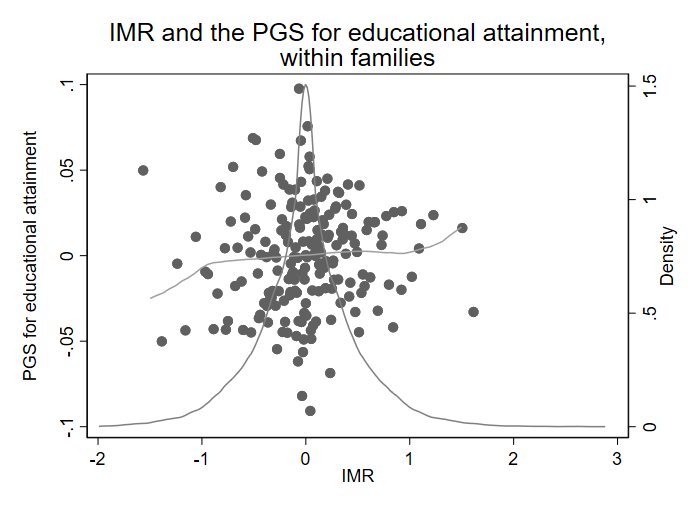} \\
  \includegraphics[width=0.45\linewidth]{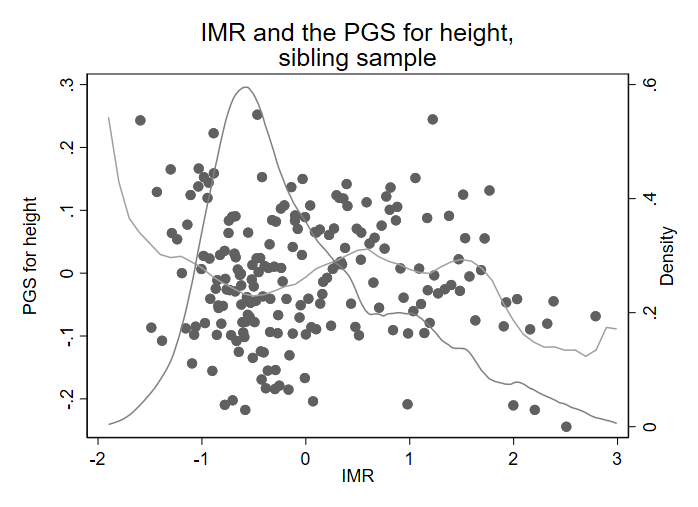}
  \includegraphics[width=0.45\linewidth]{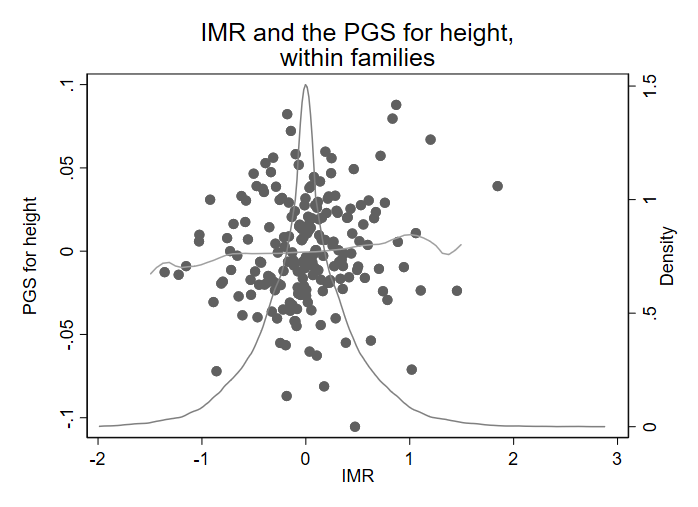} \\
  \caption{Correlation between infant mortality rates, and the polygenic scores for BMI, educational attainment and height in the sibling sample without (column 1) and with (column 2) family fixed effects.}
\label{fig:RobustnessPGS}
\end{figure}

\begin{figure}[ht]
  \centering
  \includegraphics[width=0.45\linewidth]{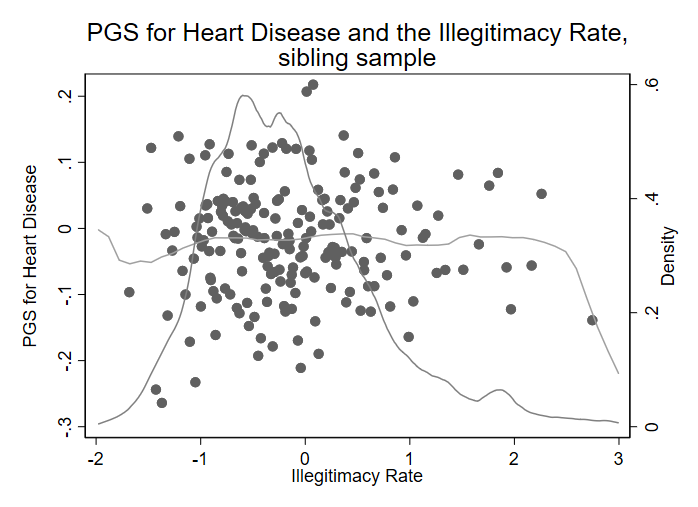}
  \includegraphics[width=0.45\linewidth]{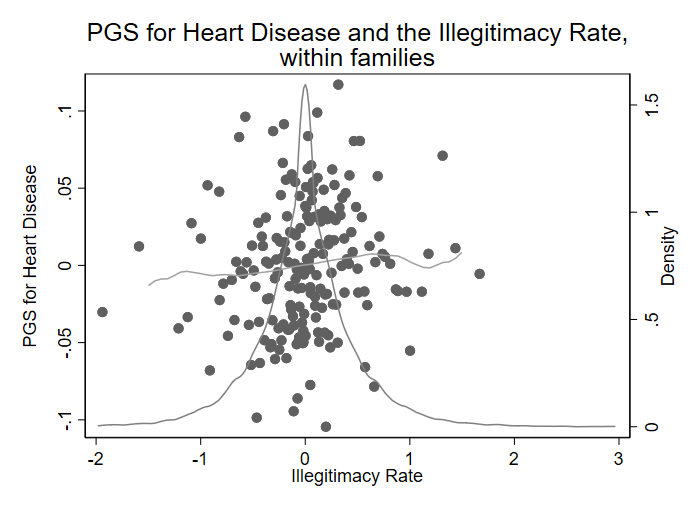} \\
  \includegraphics[width=0.45\linewidth]{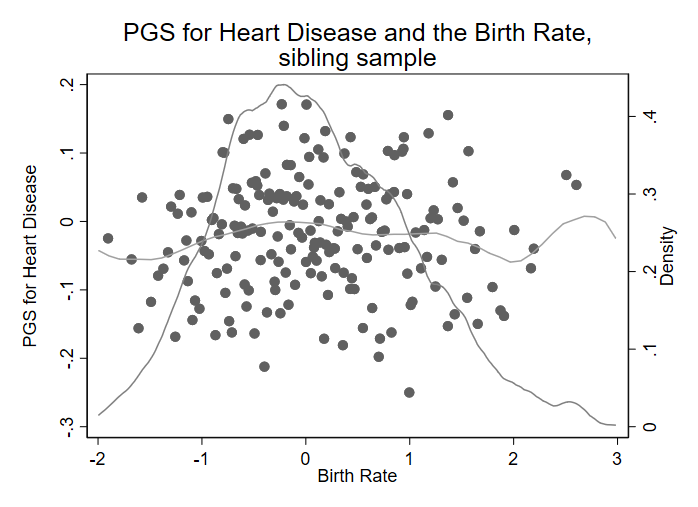} \includegraphics[width=0.45\linewidth]{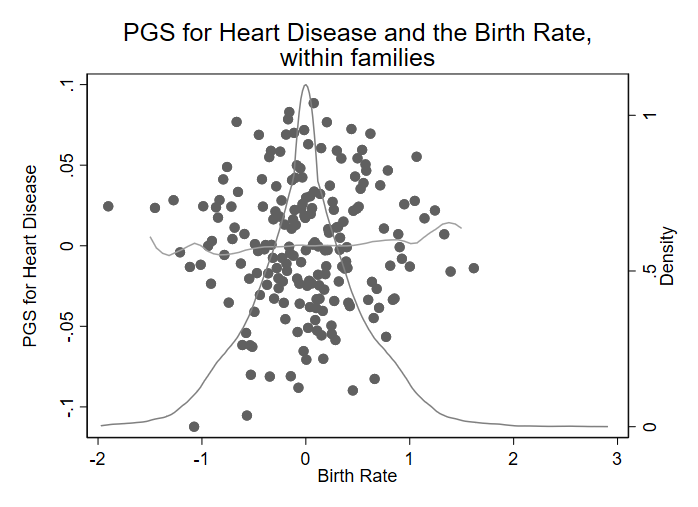} \\
  \includegraphics[width=0.45\linewidth]{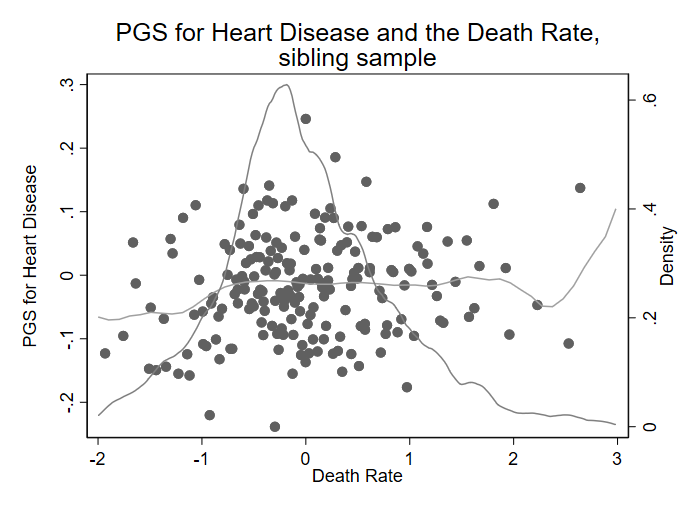}
  \includegraphics[width=0.45\linewidth]{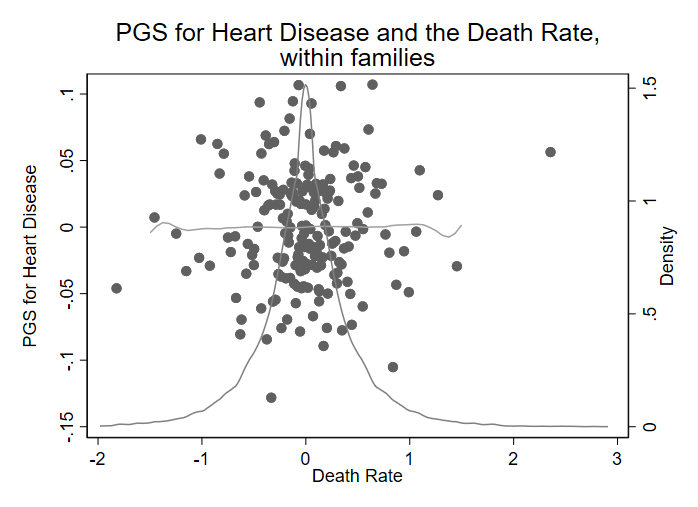} \\
  \caption{Correlation between the polygenic score for heart disease and three alternative early life environments: the district-level illegitimacy rate, birth rate and death rate  using the sibling sample without (column 1) and with (column 2) family fixed effects.}
\label{fig:RobustnessE}
\end{figure}

To further explore whether gene-environment correlation ($rGE$) may be driving our results, \autoref{fig:RobustnessPGS} presents the same (between-family and within-family) figures as \autoref{fig:DensityPGS}, focusing on the sibling sample, but instead of using the polygenic score for heart disease, we use polygenic scores that are relevant to \textit{other} traits. In \autoref{fig:RobustnessE}, we show the correlation between the polygenic score for heart disease and \textit{alternative} early life environments. Indeed, if the environment of interest (the infant mortality rate), is somehow correlated to one's \textit{other} polygenic score, or if one's polygenic score for heart disease is somehow correlated to \textit{other} early life environments, it is unclear whether the environment-coefficient partially captures one's genetic predisposition, and vice versa, through $rGE$. In that case, it is unclear whether the coefficient on the gene-environment interaction ($G \times E$) is truly a $G \times E$, or whether it captures a $G \times G$ or $E \times E$. 

For example, if individuals with a high polygenic score for educational attainment are more likely to select into healthier and wealthier environments \citep[see e.g.][]{Abdellaoui2019}, we may see a negative correlation between the polygenic score for educational attainment and the infant mortality rate. It is then unclear whether the $G \times E$ effect in the main analysis captures a true interaction effect between the polygenic score for heart disease and the early life infant mortality rate, or whether this is partially capturing an interaction between the the polygenic score for heart disease and the polygenic score for educational attainment ($G \times G$). Similarly, if individuals with higher polygenic scores for heart disease are more likely to live in areas characterised by higher illegitimacy rates, it is unclear whether the interaction effect captures a true $G \times E$ effect, or whether this picks up an interaction between the infant mortality rate and the illegitimacy rate ($E \times E$).

As genetic variants are randomly allocated within families, we do not encounter this interpretational issue with the coefficient on the polygenic score in the \textit{within-family} analysis. Indeed, the coefficient on the genetic term in the within-family analysis captures a genetic effect\footnote{Though note that the genetic effect is potentially attenuated; see \cite{Trejo2019}.} In the \textit{between-family} analysis, it may capture either a genetic or environmental effect; the latter via genetic nurture or $rGE$. 

\autoref{fig:RobustnessPGS} shows the gene-environment correlations between the infant mortality rate at birth and the polygenic scores for BMI \citep[using GWAS summary statistics from][]{Locke2015}, educational attainment (23andMe), and height \citep{Locke2015}. The left panel of each row of graphs shows the raw correlation between the infant mortality rate and the relevant polygenic score for the sibling sample. The right panel shows this same relationship exploiting only \textit{within-family} variation, i.e., taking the residuals of the variable shown on the axes after taking out the family fixed effect.

\autoref{fig:RobustnessE} shows the gene-environment correlations between the polygenic score for heart disease and alternative district-level variables: the district-level illegitimacy rate, birth rate and death rate. The left panel again shows the raw correlation for the sibling sample, whilst the panel on the right only exploits within-family variation.

For both \autoref{fig:RobustnessPGS} and \autoref{fig:RobustnessE}, the slope of the lines are close to zero. In other words, we find no evidence of any strong gene-environment correlation for any of the polygenic scores and early life environments used here. This again is reassuring, suggesting we are identifying true $G \times E$ interactions, rather than spurious $G \times G$ or $E \times E$ interactions.

\end{document}

%% file: tab/IMR_SES.tex
            &\multicolumn{1}{c}{All years}&\multicolumn{1}{c}{1951}&\multicolumn{1}{c}{1961}&\multicolumn{1}{c}{1971}\\
\midrule
Illegitimacy rate&      0.0105         &      0.0050         &     -0.0048         &      0.0303\sym{***}\\
            &    (0.0091)         &    (0.0238)         &    (0.0145)         &    (0.0106)         \\
\addlinespace
Proportion social class I&     -0.3373\sym{***}&     -0.7362\sym{***}&     -0.4171\sym{**} &     -0.2087         \\
            &    (0.1240)         &    (0.2654)         &    (0.2100)         &    (0.1519)         \\
\addlinespace
Proportion social class II &     -0.0298         &      0.1144         &     -0.0626         &     -0.0826         \\
            &    (0.0375)         &    (0.0875)         &    (0.0499)         &    (0.0629)         \\
\addlinespace
Proportion social class IV&      0.0996\sym{**} &      0.1783\sym{***}&      0.0584         &     -0.1452\sym{**} \\
            &    (0.0421)         &    (0.0650)         &    (0.0693)         &    (0.0725)         \\
\addlinespace
Proportion social class V&      0.2447\sym{***}&      0.3718\sym{***}&      0.2311         &      0.1053         \\
            &    (0.0750)         &    (0.1142)         &    (0.1478)         &    (0.1369)         \\
\midrule
R$^2$       &        0.17         &        0.03         &        0.03         &        0.03         \\
No. of observations&        4297         &        1468         &        1409         &        1420         \\

%% file: tab/IHD.tex
                    &\multicolumn{5}{c}{Ischaemic Heart Disease}                                                                  \\\cmidrule(lr){2-6}
                    &\multicolumn{1}{c}{(1)}&\multicolumn{1}{c}{(2)}&\multicolumn{1}{c}{(3)}&\multicolumn{1}{c}{(4)}&\multicolumn{1}{c}{(5)}\\
\midrule
IMR                 &      0.0156\sym{***}&                     &      0.0128\sym{***}&      0.0130\sym{***}&      0.0049\sym{***}\\
                    &    (0.0017)         &                     &    (0.0014)         &    (0.0014)         &    (0.0011)         \\
\addlinespace
IMR$^2$             &     -0.0017\sym{***}&                     &     -0.0013\sym{***}&     -0.0015\sym{***}&     -0.0005         \\
                    &    (0.0005)         &                     &    (0.0004)         &    (0.0005)         &    (0.0005)         \\
\addlinespace
PGS                 &                     &      0.0290\sym{***}&      0.0289\sym{***}&      0.0288\sym{***}&      0.0285\sym{***}\\
                    &                     &    (0.0006)         &    (0.0006)         &    (0.0006)         &    (0.0006)         \\
\addlinespace
PGS$^2$             &                     &      0.0042\sym{***}&      0.0042\sym{***}&      0.0040\sym{***}&      0.0040\sym{***}\\
                    &                     &    (0.0004)         &    (0.0004)         &    (0.0004)         &    (0.0004)         \\
\addlinespace
IMR $\times$ PGS    &                     &                     &                     &      0.0105\sym{***}&      0.0106\sym{***}\\
                    &                     &                     &                     &    (0.0006)         &    (0.0007)         \\
\addlinespace
IMR$^2 \times$ PGS$^2$&                     &                     &                     &      0.0002         &      0.0002         \\
                    &                     &                     &                     &    (0.0002)         &    (0.0002)         \\
\midrule
Covariates          &         Yes         &         Yes         &         Yes         &         Yes         &         Yes         \\
District FEs        &          No         &          No         &          No         &          No         &         Yes         \\
Mean                &        0.11         &        0.11         &        0.11         &        0.11         &        0.11         \\
R$^2$               &        0.07         &        0.07         &        0.07         &        0.08         &        0.08         \\
No. of observations &      378838         &      378838         &      378838         &      378838         &      378838         \\

%% file: tab/IHD_Sibs_noDFE.tex
                    &\multicolumn{3}{c}{Full sample}                                  &\multicolumn{3}{c}{Reduced (sibling) sample}                     &\multicolumn{3}{c}{With family fixed effects}                    \\\cmidrule(lr){2-4}\cmidrule(lr){5-7}\cmidrule(lr){8-10}
\midrule
IMR                 &      0.0136\sym{***}&      0.0128\sym{***}&      0.0130\sym{***}&      0.0129\sym{***}&      0.0121\sym{***}&      0.0121\sym{***}&     -0.0003         &     -0.0002         &     -0.0002         \\
                    &    (0.0014)         &    (0.0014)         &    (0.0014)         &    (0.0034)         &    (0.0033)         &    (0.0034)         &    (0.0049)         &    (0.0049)         &    (0.0049)         \\
\addlinespace
IMR$^2$             &     -0.0015\sym{***}&     -0.0013\sym{***}&     -0.0015\sym{***}&     -0.0012         &     -0.0011         &     -0.0016         &      0.0009         &      0.0007         &      0.0003         \\
                    &    (0.0004)         &    (0.0004)         &    (0.0005)         &    (0.0014)         &    (0.0014)         &    (0.0014)         &    (0.0023)         &    (0.0023)         &    (0.0023)         \\
\addlinespace
PGS                 &                     &      0.0289\sym{***}&      0.0288\sym{***}&                     &      0.0277\sym{***}&      0.0274\sym{***}&                     &      0.0222\sym{***}&      0.0219\sym{***}\\
                    &                     &    (0.0006)         &    (0.0006)         &                     &    (0.0017)         &    (0.0018)         &                     &    (0.0033)         &    (0.0033)         \\
\addlinespace
PGS$^2$             &                     &      0.0042\sym{***}&      0.0040\sym{***}&                     &      0.0060\sym{***}&      0.0055\sym{***}&                     &      0.0058\sym{***}&      0.0053\sym{**} \\
                    &                     &    (0.0004)         &    (0.0004)         &                     &    (0.0013)         &    (0.0014)         &                     &    (0.0019)         &    (0.0022)         \\
\addlinespace
IMR $\times$ PGS    &                     &                     &      0.0105\sym{***}&                     &                     &      0.0054\sym{***}&                     &                     &      0.0052\sym{**} \\
                    &                     &                     &    (0.0006)         &                     &                     &    (0.0018)         &                     &                     &    (0.0026)         \\
\addlinespace
IMR$^2 \times$ PGS$^2$&                     &                     &      0.0002         &                     &                     &      0.0006         &                     &                     &      0.0006         \\
                    &                     &                     &    (0.0002)         &                     &                     &    (0.0010)         &                     &                     &    (0.0012)         \\
\midrule
Controls            &         Yes         &         Yes         &         Yes         &         Yes         &         Yes         &         Yes         &         Yes         &         Yes         &         Yes         \\
District FE         &          No         &          No         &          No         &          No         &          No         &          No         &          No         &          No         &          No         \\
R$^2$               &        0.07         &        0.07         &        0.08         &        0.07         &        0.07         &        0.07         &        0.56         &        0.57         &        0.57         \\
N                   &      378838         &      378838         &      378838         &       33060         &       33060         &       33060         &       33060         &       33060         &       33060         \\

%% file: tab/mediators.tex
                    &\multicolumn{2}{c}{BMI}                    &\multicolumn{2}{c}{DBP}                    &\multicolumn{2}{c}{SBP}                    &\multicolumn{2}{c}{Height}                 &\multicolumn{2}{c}{Alcohol}                &\multicolumn{2}{c}{Smoking}                \\\cmidrule(lr){2-3}\cmidrule(lr){4-5}\cmidrule(lr){6-7}\cmidrule(lr){8-9}\cmidrule(lr){10-11}\cmidrule(lr){12-13}
                    &\multicolumn{1}{c}{(1)}&\multicolumn{1}{c}{(2)}&\multicolumn{1}{c}{(1)}&\multicolumn{1}{c}{(2)}&\multicolumn{1}{c}{(1)}&\multicolumn{1}{c}{(2)}&\multicolumn{1}{c}{(1)}&\multicolumn{1}{c}{(2)}&\multicolumn{1}{c}{(1)}&\multicolumn{1}{c}{(2)}&\multicolumn{1}{c}{(1)}&\multicolumn{1}{c}{(2)}\\
\midrule
IMR                 &      0.3431\sym{***}&      0.0744         &      0.1816         &     -0.1362         &      0.5996\sym{**} &     -0.0987         &     -0.7621\sym{***}&     -0.0110         &      0.0005         &     -0.0000         &     -0.0041         &      0.0053         \\
                    &    (0.0532)         &    (0.0668)         &    (0.1494)         &    (0.1559)         &    (0.2686)         &    (0.2833)         &    (0.0846)         &    (0.0759)         &    (0.0021)         &    (0.0029)         &    (0.0054)         &    (0.0079)         \\
\addlinespace
IMR$^2$             &     -0.0772\sym{***}&     -0.0381\sym{*}  &     -0.0967\sym{**} &     -0.0620         &     -0.2257\sym{***}&     -0.1796\sym{*}  &      0.0895\sym{**} &     -0.0217         &     -0.0022         &     -0.0030\sym{*}  &     -0.0015         &     -0.0042         \\
                    &    (0.0249)         &    (0.0223)         &    (0.0455)         &    (0.0556)         &    (0.0818)         &    (0.0971)         &    (0.0443)         &    (0.0265)         &    (0.0016)         &    (0.0016)         &    (0.0024)         &    (0.0033)         \\
\addlinespace
PGS                 &      0.3094\sym{***}&      0.2749\sym{***}&      0.5100\sym{***}&      0.4555\sym{***}&      1.0704\sym{***}&      0.9432\sym{***}&     -0.3949\sym{***}&     -0.1015\sym{**} &     -0.0031\sym{***}&     -0.0040\sym{**} &      0.0140\sym{***}&      0.0103\sym{**} \\
                    &    (0.0277)         &    (0.0399)         &    (0.0528)         &    (0.0924)         &    (0.1116)         &    (0.1855)         &    (0.0385)         &    (0.0437)         &    (0.0011)         &    (0.0020)         &    (0.0029)         &    (0.0048)         \\
\addlinespace
PGS$^2$             &     -0.0179         &     -0.0097         &     -0.0883\sym{*}  &     -0.0866         &     -0.1663\sym{*}  &     -0.1355         &      0.0419         &      0.0261         &     -0.0011         &     -0.0014         &     -0.0012         &     -0.0048\sym{*}  \\
                    &    (0.0207)         &    (0.0255)         &    (0.0486)         &    (0.0633)         &    (0.0978)         &    (0.1327)         &    (0.0309)         &    (0.0343)         &    (0.0010)         &    (0.0012)         &    (0.0023)         &    (0.0028)         \\
\addlinespace
IMR $\times$ PGS    &     -0.0604\sym{**} &     -0.0225         &     -0.1336\sym{***}&     -0.0591         &     -0.0135         &      0.0709         &     -0.0188         &     -0.0516         &      0.0006         &     -0.0013         &     -0.0038         &     -0.0001         \\
                    &    (0.0276)         &    (0.0346)         &    (0.0509)         &    (0.0807)         &    (0.1055)         &    (0.1434)         &    (0.0337)         &    (0.0411)         &    (0.0013)         &    (0.0019)         &    (0.0033)         &    (0.0041)         \\
\addlinespace
IMR$^2 \times$ PGS$^2$&      0.0104         &      0.0041         &      0.0078         &      0.0289         &      0.0642         &      0.0771         &     -0.0026         &     -0.0258         &      0.0003         &      0.0014\sym{**} &      0.0000         &      0.0024\sym{*}  \\
                    &    (0.0088)         &    (0.0131)         &    (0.0227)         &    (0.0303)         &    (0.0435)         &    (0.0584)         &    (0.0163)         &    (0.0159)         &    (0.0007)         &    (0.0007)         &    (0.0012)         &    (0.0014)         \\
\midrule
Covariates          &         Yes         &         Yes         &         Yes         &         Yes         &         Yes         &         Yes         &         Yes         &         Yes         &         Yes         &         Yes         &         Yes         &         Yes         \\
Family FEs          &          No         &         Yes         &          No         &         Yes         &          No         &         Yes         &          No         &         Yes         &          No         &         Yes         &          No         &         Yes         \\
Mean                &       27.29         &       27.29         &       82.10         &       82.10         &      137.98         &      137.98         &      168.11         &      168.11         &        0.96         &        0.96         &        0.45         &        0.45         \\
R$^2$               &        0.03         &        0.64         &        0.06         &        0.60         &        0.13         &        0.63         &        0.54         &        0.89         &        0.03         &        0.56         &        0.03         &        0.60         \\
No. of observations &       33278         &       33278         &       31842         &       31842         &       31842         &       31842         &       31842         &       31842         &       31799         &       31799         &       31594         &       31594         \\

%% file: tab/IHDgender.tex
                    &\multicolumn{5}{c}{Females}                                                                                  &\multicolumn{5}{c}{Males}                                                                                    \\\cmidrule(lr){2-6}\cmidrule(lr){7-11}
                    &\multicolumn{1}{c}{(1)}&\multicolumn{1}{c}{(2)}&\multicolumn{1}{c}{(3)}&\multicolumn{1}{c}{(4)}&\multicolumn{1}{c}{(5)}&\multicolumn{1}{c}{(6)}&\multicolumn{1}{c}{(7)}&\multicolumn{1}{c}{(8)}&\multicolumn{1}{c}{(9)}&\multicolumn{1}{c}{(10)}\\
\midrule
IMR                 &      0.0109\sym{***}&                     &      0.0109\sym{***}&      0.0110\sym{***}&      0.0043\sym{***}&      0.0151\sym{***}&                     &      0.0151\sym{***}&      0.0154\sym{***}&      0.0051\sym{**} \\
                    &    (0.0014)         &                     &    (0.0014)         &    (0.0014)         &    (0.0011)         &    (0.0018)         &                     &    (0.0018)         &    (0.0018)         &    (0.0020)         \\
\addlinespace
IMR$^2$             &     -0.0006         &                     &     -0.0006         &     -0.0012\sym{**} &     -0.0005         &     -0.0020\sym{**} &                     &     -0.0020\sym{**} &     -0.0016\sym{*}  &     -0.0001         \\
                    &    (0.0005)         &                     &    (0.0005)         &    (0.0006)         &    (0.0007)         &    (0.0008)         &                     &    (0.0008)         &    (0.0009)         &    (0.0008)         \\
\addlinespace
PGS                 &                     &      0.0170\sym{***}&      0.0170\sym{***}&      0.0171\sym{***}&      0.0169\sym{***}&                     &      0.0429\sym{***}&      0.0429\sym{***}&      0.0423\sym{***}&      0.0419\sym{***}\\
                    &                     &    (0.0005)         &    (0.0005)         &    (0.0005)         &    (0.0005)         &                     &    (0.0010)         &    (0.0010)         &    (0.0010)         &    (0.0010)         \\
\addlinespace
PGS$^2$             &                     &      0.0028\sym{***}&      0.0028\sym{***}&      0.0022\sym{***}&      0.0022\sym{***}&                     &      0.0058\sym{***}&      0.0058\sym{***}&      0.0062\sym{***}&      0.0063\sym{***}\\
                    &                     &    (0.0004)         &    (0.0004)         &    (0.0005)         &    (0.0005)         &                     &    (0.0007)         &    (0.0007)         &    (0.0007)         &    (0.0007)         \\
\addlinespace
IMR $\times$ PGS    &                     &                     &                     &      0.0075\sym{***}&      0.0076\sym{***}&                     &                     &                     &      0.0136\sym{***}&      0.0138\sym{***}\\
                    &                     &                     &                     &    (0.0006)         &    (0.0006)         &                     &                     &                     &    (0.0011)         &    (0.0011)         \\
\addlinespace
IMR$^2 \times$ PGS$^2$&                     &                     &                     &      0.0006\sym{**} &      0.0006\sym{**} &                     &                     &                     &     -0.0004         &     -0.0005         \\
                    &                     &                     &                     &    (0.0003)         &    (0.0003)         &                     &                     &                     &    (0.0004)         &    (0.0004)         \\
\midrule
Covariates          &         Yes         &         Yes         &         Yes         &         Yes         &         Yes         &         Yes         &         Yes         &         Yes         &         Yes         &         Yes         \\
District FEs        &          No         &          No         &          No         &          No         &         Yes         &          No         &          No         &          No         &          No         &         Yes         \\
Mean                &        0.07         &        0.07         &        0.07         &        0.07         &        0.07         &        0.16         &        0.16         &        0.16         &        0.16         &        0.16         \\
R$^2$               &        0.04         &        0.04         &        0.04         &        0.04         &        0.05         &        0.07         &        0.07         &        0.07         &        0.07         &        0.08         \\
No. of observations &      203750         &      203750         &      203750         &      203750         &      203738         &      175089         &      175089         &      175089         &      175089         &      175068         \\

%% file: tab/IHDrobustness.tex
                    &\multicolumn{4}{c}{Ischaemic Heart Disease}                                            \\\cmidrule(lr){2-5}
                    &\multicolumn{1}{c}{(1)}&\multicolumn{1}{c}{(2)}&\multicolumn{1}{c}{(3)}&\multicolumn{1}{c}{(4)}\\
\midrule
IMR                 &     -0.0002         &     -0.0008         &      0.0003         &      0.0009         \\
                    &    (0.0049)         &    (0.0050)         &    (0.0049)         &    (0.0049)         \\
\addlinespace
IMR$^2$             &      0.0003         &      0.0003         &      0.0004         &                     \\
                    &    (0.0023)         &    (0.0023)         &    (0.0023)         &                     \\
\addlinespace
PGS                 &      0.0219\sym{***}&      0.0219\sym{***}&      0.0324\sym{***}&      0.0217\sym{***}\\
                    &    (0.0033)         &    (0.0033)         &    (0.0032)         &    (0.0033)         \\
\addlinespace
PGS$^2$             &      0.0053\sym{**} &      0.0053\sym{**} &      0.0051\sym{**} &                     \\
                    &    (0.0022)         &    (0.0022)         &    (0.0021)         &                     \\
\addlinespace
IMR $\times$ PGS    &      0.0052\sym{**} &      0.0052\sym{**} &      0.0098\sym{***}&      0.0050\sym{*}  \\
                    &    (0.0026)         &    (0.0026)         &    (0.0029)         &    (0.0026)         \\
\addlinespace
IMR$^2$ $\times$ PGS$^2$&      0.0006         &      0.0006         &      0.0004         &                     \\
                    &    (0.0012)         &    (0.0012)         &    (0.0012)         &                     \\
\addlinespace
Illegitimacy Rate   &                     &     -0.0002         &                     &                     \\
                    &                     &    (0.0038)         &                     &                     \\
\addlinespace
Birth Rate          &                     &      0.0017         &                     &                     \\
                    &                     &    (0.0043)         &                     &                     \\
\addlinespace
Death Rate          &                     &      0.0020         &                     &                     \\
                    &                     &    (0.0035)         &                     &                     \\
\midrule
Covariates          &         Yes         &         Yes         &         Yes         &         Yes         \\
PCs                 &         Yes         &         Yes         &         Yes         &         Yes         \\
Additional controls &          No         &         Yes         &          No         &          No         \\
R$^2$               &        0.57         &        0.57         &        0.57         &        0.57         \\
No. of observations &       33060         &       33060         &       33060         &       33060         \\